\documentstyle[12pt,epsf]{article}
\newcommand{\be}{\begin{eqnarray}}
\newcommand{\ee}{\end{eqnarray}}
\newcommand{\half}{{\textstyle{\frac{1}{2}}}}
\newcommand{\textfrac}[2]{{\textstyle{\frac{#1}{#2}}}}

\newcommand{\fourint}[1]{\int\!\frac{d^4 #1}{(2\pi)^4}}
\newcommand{\Fdual}{\widetilde{F}}
\newcommand{\effop}[1]{\mbox{``$#1$''}} 
\newcommand{\partialright}{\stackrel{\rightarrow}{{}\partial}}
\newcommand{\partialleft}{\stackrel{\leftarrow}{{}\partial}}
\newcommand{\partialboth}{\stackrel{\leftrightarrow}{{}\partial}}

\newcommand{\partialrightsq}{\stackrel{\rightarrow}{{}\partial^2}}
\newcommand{\partialleftsq}{\stackrel{\leftarrow}{{}\partial^2}}
\renewcommand{\theequation}{\arabic{section}.\arabic{equation}}
\begin{document}
%
%
\rightline{RUB-TPII-5/99}
\rightline{hep-ph/9906444}
\vspace{.3cm}
\begin{center}
\begin{large}
{\bf Twist--4 contribution to unpolarized structure functions 
$F_L$ and $F_2$ from instantons} 
\\
\end{large}
\vspace{1.4cm}
{\bf B. Dressler}$^{\rm a, 1}$,  
{\bf M. Maul}$^{\rm b, 2}$ 
{\bf and C. Weiss}$^{\rm a, 3}$  
\\[1.cm]
$^{a}$ {\it Institut f\"ur Theoretische Physik II,
Ruhr--Universit\"at Bochum, \\ D--44780 Bochum, Germany} 
\\
{\it $^{b}$ NORDITA, Blegdamsvej 17, 2100 Copenhagen \O , Denmark}
\end{center}
\vspace{1.5cm}
\begin{abstract}
\noindent
We compute in the instanton vacuum the nucleon matrix elements of the
twist--4 QCD operators describing power corrections to the second moments
of the unpolarized structure functions, $F_L$ and $F_2$. Our approach takes
into account the leading contribution in the packing fraction of the
instanton medium, $\bar\rho / \bar R \ll 1$. Parametrically leading are the
matrix elements of a twist--4 quark--gluon operator, which are of the order
of the inverse instanton size, $1/\bar\rho^2 = (600\, {\rm MeV})^2$. The
matrix elements of the four--fermion (diquark) operators are suppressed by
a factor $(\bar\rho / \bar R)^4$ and numerically small. These results are
in agreement with the pattern of phenomenological $1/Q^2$--corrections to
$R = \sigma_L / \sigma_T$ and $F_2$ found in QCD fits to the data.  In
particular, the rise of $R$ at low $Q^2$ can be obtained from
instanton--type vacuum fluctuations at a low scale.
\end{abstract}
\vspace{1cm}
\vfill
\rule{5cm}{.15mm}
\\
\noindent
{\footnotesize $^{\rm 1}$ E-mail: birgitd@tp2.ruhr-uni-bochum.de} \\
{\footnotesize $^{\rm 2}$ E-mail: maul@nordita.dk} \\
{\footnotesize $^{\rm 3}$ E-mail: weiss@tp2.ruhr-uni-bochum.de}
%
%
\newpage 
\tableofcontents
\newpage
\section{Introduction}
\setcounter{equation}{0}
An outstanding problem in the theory of deep--inelastic scattering is to
understand the transition from the asymptotic region, where the scale
dependence of structure functions is described by perturbative evolution,
to the region of low $Q^2$, where non-perturbative effects play an
essential role. Starting from high $Q^2$, the onset of non-perturbative
effects manifests itself in power ($1/Q^2$--) corrections, which are
governed by non-perturbative scale parameters.  Aside from the
well--understood target mass corrections there are dynamical power
corrections, which in QCD are determined by nucleon matrix elements of
operators of non--leading twist $(>2)$ \cite{SV82,JS82,EFP82}. In the
partonic language, they describe the effects of the interaction of the
quarks with the non-perturbative gluon field in the nucleon, or of 2--body
correlations between quarks, on the DIS cross section. A well--known
example are the longitudinal structure function $F_L = F_2 - 2 x F_1$, or
the ratio $R = \sigma_L / \sigma_T$, which both happen to be zero in the
naive parton model ({\it i.e.}, without radiative corrections). QCD fits of
the data, using existing NLO parametrizations to describe the twist--2
contribution, leave room for sizable $1/Q^2$--power corrections
\cite{Whitlow90,Sanchez91,Bodek98,RSB99}. It remains a challenge to
understand the origin of these corrections from first principles.
\par
During the last years there has been mounting evidence for an important
role of instanton--type vacuum fluctuations in determining the properties
of the low--energy ha\-dro\-nic world \cite{SchSh96}. The so-called
instanton vacuum is a variational approximation to the ground state of
Yang--Mills theory in terms of a ``medium'' of instantons and
antiinstantons, with quantum fluctuations about them, which is stabilized
by instanton interactions \cite{DP84}. The strong coupling constant is
fixed at a scale of the order of the inverse average size of the instantons
in the medium, $\bar\rho^{-1} \simeq 600\, {\rm MeV}$. The most important
property of this picture, which had been postulated on phenomenological
grounds by Shuryak \cite{Shuryak82} and was later established in the
variational calculation of Diakonov and Petrov \cite{DP84}, is the
diluteness of the instanton medium, {\it i.e.}, the fact that the ratio of
the average size of instantons in the medium to their average distance,
$\bar R$, is small: $\bar\rho / \bar R \simeq 1/3$. The existence of this
small parameter makes possible a systematic analysis of non-perturbative
effects in this approach.
\par
In particular, the instanton vacuum explains the dynamical breaking of
chiral symmetry, which is crucial in determining the structure of strong
interactions at low energies \cite{DP86,DP86_prep}.  It happens due to the
delocalization of the fermionic zero modes associated with the individual
(anti--) instantons in the medium. Using the $1/N_c$--expansion one derives
from the instanton vacuum an effective low--energy theory, whose degrees of
freedom are pions (Goldstone bosons) and massive ``constituent''
quarks. The nucleon is described as a chiral soliton in this effective
theory, as a state of $N_c$ ``valence'' quarks bound by a classical pion
field \cite{DPP88}.
\par
The instanton vacuum has also received strong support from direct lattice
simulations. Not only were the basic characteristics of the instanton
medium reproduced after ``cooling'' of the quantum fluctuations
\cite{cool}, also the correlation functions of various hadronic currents
were found to remain practically unchanged under cooling, showing that the
instanton vacuum takes into account the relevant non-perturbative
fluctuations in these channels.
\par
Given the general success of the instanton vacuum in describing hadronic
properties, it is natural to ask if this scheme of approximations can
explain the non-perturbative effects observed in the scale dependence of
structure functions at low $Q^2$. More ambitiously, one could also ask the
opposite question: Can we use the rich and increasingly accurate
information obtained from DIS to learn something about properties of
non-perturbative vacuum fluctuations?
\par
The aim of this paper is to study the $1/Q^2$--power corrections to
unpolarized structure functions, in particular, to the longitudinal
structure function, $F_L$, in the instanton vacuum. We restrict ourselves
to the lowest moments, for which the power corrections can be studied using
the operator product expansion (OPE) \cite{SV82,JS82}.  In particular, we
use the instanton vacuum to compute the nucleon matrix elements of the
twist--4 operators, both the quark--gluon and the four--quark operators,
and discuss their relative magnitude. Our approach takes into account the
leading contribution in the packing fraction of the instanton medium,
$\bar\rho / \bar R$ \cite{DPW96,BPW97}.  Our main conclusion is that the
matrix element of the twist--4 quark--gluon correlator is parametrically
large and accounts for most of the observed scaling violations in $F_L$,
while the four--fermionic operators are suppressed.
\par
The quantitative description of power corrections is a complex problem. The
set of higher--twist operators used in the OPE is over complete, since
matrix elements of different operators are related by the QCD equations of
motion.  In order to have results which are independent of the choice of
the operator basis one needs to make consistent approximations in dealing
with the QCD operators and with the nucleon state, something which is
generally impossible to achieve in phenomenological approaches such as the
bag model \cite{JS82}. In this respect the instanton vacuum with its
inherent small parameter, $\bar\rho / \bar R$, offers some decisive
methodological advantages. As was shown in Ref.\cite{BPW97}, in this
approach the QCD equations of motion are preserved at the level of
effective operators, which is a consequence of the one--instanton
approximation justified in leading order of $\bar\rho / \bar R$, and of the
fact that the fermions couple to the instanton through the zero mode, see
\cite{BPW97} for details. It is crucial in this context that the
description of the nucleon as a chiral soliton is a fully field--theoretic
one and does not involve any additional dynamical input besides the chiral
symmetry breaking induced by the instanton fluctuations.  Also, it should
be pointed out that, contrary to other field--theoretic approaches such as
QCD sum rules or lattice calculations our approach is based on a small
parameter, $\bar\rho / \bar R$, which endows the {\it a priori}
structure-less set of higher--twist matrix elements with a parametric
order, thus greatly simplifying the discussion of power corrections.
\par
The plan of this paper is as follows. In Section \ref{sec_power}, we
summarize the QCD predictions for the scale dependence of the unpolarized
structure functions $F_L$ and $F_2$ to order $1/Q^2$, including twist--2,
target mass and dynamical twist--4 corrections. The central part of the
paper, Section \ref{sec_instantons}, is devoted to the calculation of the
relevant twist--4 matrix elements in the instanton vacuum.  Subsection
\ref{subsec_instanton_vacuum} describes the basic properties of the
instanton vacuum and the resulting low-energy effective chiral theory. In
Subsection \ref{subsec_effective_operators} we outline the method by which
the QCD twist--4 operators can be mapped onto effective operators, whose
matrix elements can be computed in the low-energy effective theory
\cite{BPW97}. As a novel feature, we show here that the QCD twist--4
quark--gluon operators which produce parametrically large matrix elements
in $\bar\rho / \bar R$ can be represented in the effective theory by
one--body operators built from the massive ``constituent'' quark
field. This representation turns out to be very convenient both for
discussing general properties of instanton--induced higher--twist effects,
as well as for computing the nucleon matrix elements. In Subsection
\ref{subsec_nucleon} we compute the nucleon matrix elements of the
effective operators, within the large--$N_c$ picture of the nucleon as a
chiral soliton. In Section \ref{sec_comparison} we use the instanton
results for the twist--4 matrix elements to estimate the scale dependence
of $F_L$ and $F_2$, and compare with the data for $R = \sigma_L / \sigma_T$
and $F_2$.  We also compare our results for the matrix elements with those
of other theoretical approaches. Conclusions and an outlook are presented
in Section \ref{sec_conclusions}.
\par
Appendix \ref{app_ope} contains a short derivation of the QCD expressions
for the dynamical twist--4 corrections to unpolarized structure functions
using the operator product expansion \cite{SV82}.  Since calculations of
higher--twist effects are highly dependent on conventions we explicitly
state all relevant conventions there.
\par
We are not considering here power corrections to DIS structure functions
resulting from instanton contributions to the coefficient functions of the
OPE, {\it i.e.}, semiclassical corrections associated with single
instantons of size $\rho \sim 1/Q$ \cite{BB93,Moch96}. These are suppressed
by very large powers of $1/Q$ and typically negligibly small. Rather, we
use a medium of instantons of average size 
$\bar\rho = (600\, {\rm MeV})^{-1}$ as a model for the non-perturbative
vacuum structure at the hadronic scale, in order to calculate the matrix
elements parametrizing $1/Q^2$--corrections.
\par
The power corrections to $F_L$ and $R = \sigma_L / \sigma_T$ have been
estimated previously in a variety of models. In the so-called transverse
basis of twist--4 operators the $1/Q^2$--corrections to $F_L$ can be
related to an ``intrinsic'' transverse momentum of the partons
\cite{EFP82}. This language is frequently used to model dynamical twist--4
contributions. A model interpolating between large $Q^2$ and the limit $Q^2
\rightarrow 0$ has been proposed in Ref.\cite{BKS97}. In Ref.\cite{Guo98}
the leading--order twist--4 contribution at large $x$ has been related to
the derivative of the twist--2 distribution. Also, the magnitude of the
twist--4 contribution to $F_L$ has been estimated from the renormalon
ambiguity of the perturbation series for the $O(\alpha_S )$ twist--2
coefficient functions calculated in the large--$N_f$ limit
\cite{Stein96,Stein98}.  In Ref.\cite{YML98} an attempt was made to model
the non-perturbative twist--4 contribution at large $x$ by introducing a
coupling to vacuum condensates. Higher--twist contributions to unpolarized
structure functions have also been studied in the bag model
\cite{Choi93,Signal97}.
\section{Power corrections to unpolarized DIS}
\setcounter{equation}{0}
\label{sec_power}
The hadronic tensor for unpolarized electromagnetic scattering off the
nucleon has two independent symmetric structures, which are parameterized
in terms of the two structure functions $F_1$ and $F_2$, or, equivalently,
$F_L \equiv F_2 - 2 x F_1$ and $F_2$:
\be
W_{\mu\nu} (p, q) &\equiv& \frac{1}{4\pi} \int d^4 z \; e^{i q\cdot z}
\langle p | \,\, J_\mu^\dagger (z) \, J_\nu (0) \,\,
| p \rangle 
\nonumber \\
&=& \frac{F_L (x, Q^2 )}{2 x} 
\left( g_{\mu\nu} - \frac{q_\mu q_\nu}{q^2} \right)
\nonumber \\
&+& \; \frac{F_2 (x, Q^2 )}{2 x} 
\left[ \frac{-q^2}{(p\cdot q)^2} p_\mu p_\nu 
+ \frac{1}{p\cdot q} (p_\mu q_\nu + p_\nu q_\mu ) 
 - g_{\mu\nu} \right] .
\label{W_def}
\ee
Here, $J_\mu (z) = J_\mu^\dagger (z)$ is the electromagnetic current
operator, $p_\mu$ the nucleon momentum ($p^2 = M_N^2$), 
$x \equiv -q^2/(2 p\cdot q )$ and $Q^2 \equiv -q^2$. Here and in the
following, all matrix elements are understood to be averaged over the
target spin,
\be
\langle p | \ldots | p \rangle &\equiv&
\half \sum_S \langle p S | \ldots | p S \rangle ,
\ee
and the states are normalized according to $\langle p' | p \rangle = 2 p^0
(2\pi )^3 \delta^{(3)} ({\bf p}' - {\bf p})$. In connection with
experiments one frequently uses a slightly modified definition of the
longitudinal structure function \cite{Sanchez91}:
\be
\widetilde{F}_L (x, Q^2) &\equiv& F_2 (x, Q^2) \left( 1 + 
\frac{4 M_N^2 x^2}{Q^2} \right) - 2 x F_1 (x, Q^2) 
\nonumber \\
&=& F_L (x, Q^2) + \frac{4 M_N^2 x^2}{Q^2} F_2 (x, Q^2) .
\label{F_L_ex}
\ee
This function is directly related to the experimentally measured ratio 
$R = \sigma_L / \sigma_T$, namely
\be
\widetilde{F}_L (x, Q^2) &=& F_2(x, Q^2)
\left(1 + \frac{4 M_N^2 x^2}{Q^2} \right) 
\frac{R(x, Q^2)}{R(x, Q^2) + 1} .
\label{F_L_ex_from_R}
\ee
\par
In QCD, the electromagnetic current is carried by the quark field,
\be
J_\mu (z) &=& \sum_f e_f \bar\psi_f (z) \gamma_\mu \psi_f (z) ,
\ee
where $e_f$ are the quark charges.  The scale dependence of the moments of
the structure functions up to level $1/Q^2$ can be determined by relating
the hadronic tensor to the imaginary part of the forward amplitude for the
scattering of a virtual photon off the nucleon, and applying the operator
product expansion (OPE) to the time--ordered product of electromagnetic
currents in the latter \cite{SV82,JS82}. A short summary of this approach,
including all relevant conventions, is given in Appendix \ref{app_ope}.  We
consider here the second moments of the structure functions:
\be
\left.
\begin{array}{r}
\widetilde{M}_L (2, Q^2) \\[1ex]
M_{2} (2, Q^2)
\end{array} \right\}
&\equiv& \int_0^1 dx \; 
\left\{
\begin{array}{r}
\widetilde{F}_L (x, Q^2) \\[1ex]
F_{2} (x, Q^2)
\end{array} \right. .
\label{moments_def}
\ee
We distinguish between the scaling (non--power suppressed) part of these
moments, which comes from the contribution of twist--2 spin--2 operators in
the OPE, the target mass corrections ($\propto M_N^2 / Q^2$), which arise
from twist--2 spin--4 operators, and the dynamical power corrections coming
from contributions of operators of twist 4 and spin 2.
\par
In the case of the longitudinal structure function the twist--2
contribution is of order $\alpha_S$, since $F_L$ is zero in the naive
parton model due to the Callan--Gross relation. The coefficient functions
have been computed in Ref.\cite{Sanchez91}, see also
Refs.\cite{Stein98,Zijlstra92}. In leading order\footnote{We write down the
expressions for $\widetilde{F}_L$, Eq.(\ref{F_L_ex}); the only difference
to the original $F_L$ will be in the target mass corrections.}:
\be
\widetilde{M}_L (2, Q^2 )^{\rm twist-2} &=& 
\frac{\alpha_S (Q^2)}{4\pi} 
\left[ \frac{4 C_F}{3} \sum_f e_f^2 L^{(2)}_f
\; + \; \frac{2}{3} \left( \sum_f e_f^2 \right) L^{(2)}_G 
\right] ,
\label{M_L_ex_twist2}
\\
M_2 (2, Q^2)^{\rm twist-2} &=& \sum_f e_f^2 L^{(2)}_f
\nonumber \\
&+& \frac{\alpha_S (Q^2)}{4\pi} \left[ 
\frac{C_F}{3} \sum_f e_f^2 L^{(2)}_f
\; - \; \frac{1}{2} \left( \sum_f e_f^2 \right) L^{(2)}_G 
\right] ,
\label{M_2_twist2}
\ee
where $C_F = 4/3$ and the sums run over all light flavors 
($u, d$ and $s$) in the target.
Here $L^{(n)}_f, L^{(n)}_G$ denote the matrix elements of the 
twist--2 spin--$n$ quark and gluon operators in the target,
\be
\left.
\begin{array}{rcr}
\langle p | \; \bar\psi_f \gamma_{\mu_1}
i\nabla_{\mu_2} \ldots i\nabla_{\mu_n} 
\psi_f \; | p \rangle - \mbox{traces} &=& 2 L^{(n)}_f
\\[2ex]
\langle p | \; F_{\mu_1 \alpha} D_{\mu_2} 
\ldots D_{\mu_{n - 1}} F^\alpha_{\;\;\mu_n} \; | p \rangle 
- \mbox{traces}
&=& 2 L^{(n)}_G 
\end{array}
\right\}
\times \left( p_{\mu_1} \ldots p_{\mu_n} - \mbox{traces} \right)
\label{twist2_def} ,
\ee
where on the L.H.S.\ complete symmetrization in 
$\mu_1 \ldots \mu_n$ is implied. Here
\be
\nabla_\mu &\equiv& \partial_\mu - i A_\mu^a (x) \frac{\lambda^a}{2} ,
\label{covariant_fundamental_def}
\ee 
with ${\rm tr}[\lambda^a \lambda^b] = 2 \delta^{ab}$,
is the covariant derivative in the fundamental
representation of $SU(3)$,
\be
F_{\mu\nu} &\equiv& i \left[ \nabla_\mu , \nabla_\nu \right]
\;\; = \;\; F_{\mu\nu}^a \frac{\lambda^a}{2}
\label{F_from_commutator}
\ee
the gauge field, and
\be
(D_\alpha F_{\mu\nu})^a  
&=& \partial_\alpha F_{\mu\nu}^a - f^{abc} A_\alpha^b F_{\mu\nu}^c
\ee
the covariant derivative in the adjoint representation.
\par
The target mass corrections to the second moments are proportional to the
matrix element of the spin--4 twist--2 operator (see Refs.\cite{SV82,EFP82}
and Appendix \ref{app_ope}):
\be
\widetilde{M}_L (2, Q^2 )^{\rm target\; mass} 
&=& \frac{M_N^2}{Q^2} \sum_f e_f^2 L^{(4)}_f ,
\label{M_L_ex_target} 
\\
M_2 (2, Q^2 )^{\rm target\; mass} 
&=&
\;\; \frac{M_N^2}{2 Q^2} \sum_f e_f^2 L^{(4)}_f ,
\label{M_2_target}
\ee
where we have restricted ourselves to the $O(\alpha_S^0 )$ (tree--level)
coefficient function\footnote{For the second moment of $F_L$ instead of
$\widetilde{F}_L$ the coefficient in Eq.(\ref{M_L_ex_target}) would be $-3$
instead of $+1$, {\it cf.}\ Eq.(\ref{F_L_ex}).}.
\par
\begin{figure}
\setlength{\epsfxsize}{15cm}
\setlength{\epsfysize}{7.5cm}
\epsffile{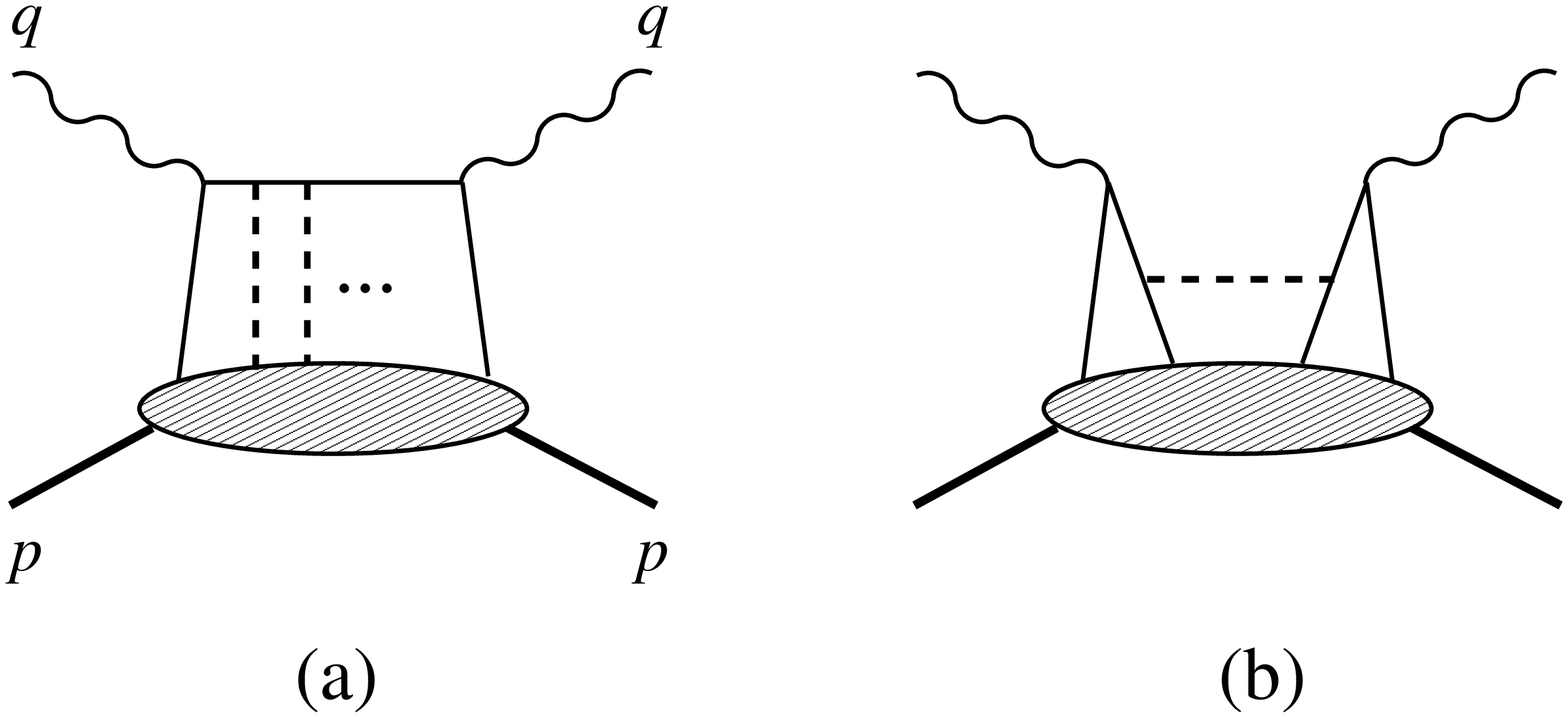}
\caption[]{The two types of twist--4 contributions to the DIS cross
section. (a) Interference between scattering of the photon off a free quark
and scattering off a quark interacting with the non-perturbative gluon
field of the nucleon. (b) Interference between scattering off two different
quarks in the nucleon, with a perturbative gluon exchange.}
\label{fig_twist4}
\end{figure}

Our main object of interest are the dynamical power corrections, which are
determined by matrix elements of twist--4 spin--2 operators
\cite{SV82,JS82}. There are two types of twist--4 contributions, shown
schematically in Fig.\ref{fig_twist4}, resulting from different
contractions of the quark fields in the two electromagnetic current
operators. Contribution (a) describes the interference of the scattering of
the photon off a free quark with scattering off a quark interacting with
the non-perturbative gluon field in the target; this contribution gives
rise to matrix elements of twist--4 quark--gluon operators in the target.
Contribution (b) describes the interference of scattering off two different
quarks in the target, accompanied by a perturbative gluon exchange; it
leads to matrix elements of four--fermionic (``diquark'') operators in the
target.  In the longitudinal structure function the latter contribution is
absent. We have recalculated the coefficient functions of the twist--4
operators, following the approach of Ref.\cite{SV82}, and reproduced the
results given in that paper.  A short outline of the calculation is given
in Appendix \ref{app_ope}.  For the second moment of the longitudinal
structure function one finds
\be
\widetilde{M}_L (2, Q^2 )^{\rm twist-4}
&=&  \frac{1}{Q^2} \sum_f e_f^2 \left(
 -\frac{1}{4} A_f + \frac{3}{4}  B_f \right) .
\label{M_L_twist4}
\ee
Again we have restricted ourselves to the tree--level coefficient function.
Here, $A_f$ and $B_f$ [their dimension is $(\mbox{mass})^2$] denote the
reduced matrix elements of the twist--4 spin--2 operators:
\be
A_{\alpha\beta, \; f} &=& \bar\psi_f \gamma_\alpha (D^\gamma 
F_{\gamma\beta}) \psi_f
\; - \; \mbox{trace} ,
\label{A_def}
\\[1ex]
B_{\alpha\beta, \; f} &=& \bar\psi_f \gamma^\gamma \gamma_5 i \left(
\nabla_{\alpha} \Fdual_{\beta\gamma } 
+ \Fdual_{\alpha\gamma} \nabla_{\beta} 
\right)
\psi_f
\; - \; \mbox{trace} ,
\label{B_def}
\ee
with
\be
\langle p| 
\left\{\begin{array}{r} 
A_{\alpha\beta, \; f} \\[1ex] B_{\alpha\beta, \; f}
\end{array}\right\} 
|p \rangle &=& 2 
\left\{\begin{array}{r} 
A_f \\[1ex] B_f
\end{array}\right\} 
\times
\left( p_\alpha p_\beta - \textfrac{1}{4} p^2 g_{\alpha\beta} \right) .
\label{me_A_B}
\ee
It is understood that the operators are symmetrized in the Lorentz indices
according to $A_{\alpha\beta} \rightarrow \frac{1}{2} ( A_{\alpha\beta} +
A_{\beta\alpha})$; this will not be explicitly indicated in the
following. Here
\be
\Fdual^{\mu\nu} &=& \frac{1}{2} \epsilon^{\mu\nu\rho\sigma} F_{\rho\sigma}
\ee 
is the dual field strength. We are using the convention 
$\gamma_5 = -i \gamma^0 \gamma^1 \gamma^2 \gamma^3$ (as in
Refs.\cite{SV82,DP86,DPW96,BPW97}), which differs from the one of Bjorken
and Drell by a minus sign, and $\epsilon^{0123} = 1$.  Note that, using the
QCD equations of motion, one can replace the forward matrix element of the
operator $A_{\alpha\beta, \; f}$ by that of a four--fermionic operator:
\be
A_{\alpha\beta, \; f}
&\rightarrow& -g^2 
\left( \bar\psi_f \frac{\lambda^a}{2} \gamma_{\alpha} 
\psi_f \right)
\left( \sum_{f'} \bar\psi_{f'} 
\frac{\lambda^a}{2} \gamma_{\beta} \psi_{f'} \right)
\label{A_fourfermion} .
\ee
\par
In the case of the structure function $F_2$ one has to consider also the
twist--4 contributions of type Fig.\ref{fig_twist4} (b), which can be
computed using standard Feynman rules. One finds
\be
M_2 (2, Q^2 )^{\rm twist-4}
&=&  
\frac{1}{Q^2} \sum_f e_f^2 \left(
 -\frac{5}{8} A_f - \frac{1}{8} B_f \right) + \frac{1}{Q^2} \sum_{f,f'} e_f
e_{f'} C_{ff'} \; ,
\label{M_2_twist4}
\ee
where $C_{ff'}$ denotes the reduced matrix element [{\it cf.}\
Eq.(\ref{me_A_B})] of the four--fermionic operator
\be
C_{\alpha\beta, \; ff'} &=& g^2 
\left( \bar\psi_f \frac{\lambda^a}{2} \gamma_\alpha\gamma_5 \psi_f \right)
\left( \bar\psi_{f'} 
\frac{\lambda^a}{2} \gamma_\beta\gamma_5 \psi_{f'} \right) .
\label{C_def}
\ee
\section{Twist--4 contribution to $F_L$ and $F_2$ from instantons}
\setcounter{equation}{0}
\label{sec_instantons}
\subsection{Instanton vacuum and low--energy effective theory}
\label{subsec_instanton_vacuum}
The framework for our calculation of matrix elements of
QCD operators is the instanton--based vacuum
of Diakonov and Petrov \cite{DP84}, which is a variational
estimate of the Euclidean Yang--Mills (``quenched QCD'') partition
function in terms of configurations consisting of a superposition
of instantons and antiinstantons ($I$'s and $\bar I$'s), with
quantum fluctuations about them. The medium of $I$'s and $\bar I$'s 
stabilizes due to an effective repulsion between the pseudoparticles.
The resulting statistical ensemble turns out to be well described by 
a partition function of independent $I$'s and $\bar I$'s with an 
effective size distribution, with average size 
$\bar\rho = (600\, {\rm MeV})^{-1}$.
In the large--$N_c$ limit the size distribution is sharply peaked, 
so one may take the sizes of all $I$'s and $\bar I$'s to be
equal to $\bar\rho$. An important property, which is in fact 
crucial for this picture to be consistent, is the diluteness of the
medium of $I$'s and $\bar I$'s. In the approach of Ref.\cite{DP84}
the ratio of the average size of instantons to their average distance 
in the medium was found to be $\bar\rho / \bar R \simeq 1/3 \ll 1$. This 
fact can be traced back to the ``accidentally'' large value of the 
coefficient of the one--loop beta function of QCD, 
$b \equiv 11 N_c/3 \approx 10$.
\par
When quarks are included, the dynamical breaking of chiral symmetry 
is a consequence of the
interaction of the quarks with the fermionic zero modes associated
with the individual instantons \cite{DP86}. The most compact way to 
express this is through an effective low--energy theory, which has been
derived in the large--$N_c$ limit using the so-called zero mode 
approximation (which is exact in leading order of $\bar \rho / \bar R$)
\cite{DP86_prep,DPW96}. It can be
formulated in terms of quark fields ($N_f$ flavors) with a dynamical mass 
and a Goldstone bosons (pion) field, with an effective 
action
\be
S_{\rm eff} &=& \int d^4 x \sum_{f,f' = 1}^{N_f}
\psi^\dagger_{f'} (x) \left[ \delta_{f'f} \; (-i) \gamma_\mu \partial_\mu
\; - \; 
i M F( -\!\partialleftsq ) \, U^{\gamma_5} (x)_{f'f} \,
F(-\!\partialrightsq ) \right] \psi_f (x) . \;\;\;
\;\;\;\;
\label{S_eff}
\ee
We have passed to Euclidean field theory by continuing the time dependence 
of the fields to imaginary times and introducing Euclidean fields and
gamma matrices according to the conventions given in 
Appendix \ref{app_eucl}. (Unless stated otherwise, all formulas in 
Subsections \ref{subsec_instanton_vacuum} and
\ref{subsec_effective_operators} will 
be written in terms of Euclidean quantities.)
In Eq.(\ref{S_eff}), $M$ is the dynamical quark mass; parametrically 
it is of order
\be
M\bar\rho &\sim & \left(\frac{\bar\rho}{\bar R}\right)^2 .
\ee
The pion field, $\pi (x)$, couples to the quarks in a chirally invariant 
way,
\be
U^{\gamma_5} (x) &=& \frac{1 + \gamma_5}{2} U (x)
+ \frac{1 - \gamma_5}{2} U^\dagger (x) ,
\hspace{1cm} U (x) \;\; = \;\; 
\exp\left[ i\pi^a (x) \tau^a \right] .
\label{U}
\ee
Furthermore, $F(-\partial^2 )$ is a form factor proportional to the
wave function of the fermionic zero mode of the instanton in momentum 
representation \cite{DP86,DPW96},
\be
F(k^2 ) &=& - t \frac{d}{dt} 
\left[ I_0 \left(\frac{t}{2}\right) K_0 \left(\frac{t}{2}\right) 
- I_1 \left(\frac{t}{2}\right) K_1 \left(\frac{t}{2}\right)
\right], \hspace{1.5cm}
t \; \equiv \; \bar\rho\sqrt{k^2} ,
\ee
where $I_n$ and $K_n$ are modified Bessel functions.
It satisfies $F(0) = 1$, and $F(k^2 ) \sim (\bar \rho^2 k^2)^{-3/2}$ for 
$k^2 \rightarrow \infty$, {\it i.e.}, $F(k^2 )$ drops to zero for momenta 
of the order of the inverse instanton size. Here and 
in the following we use a shorthand notation for the action of the form 
factors,
\be
\left.
\begin{array}{c}
F(-\!\partialrightsq ) \, \psi (x) \\[1ex]
\psi^\dagger (x) \, F(-\!\partialleftsq ) 
\end{array}
\right\}
&\equiv& \fourint{k} e^{\pm i k\cdot (x - x')} F(k^2 ) 
\left\{
\begin{array}{r}
\psi (x') \\[1ex]
\psi^\dagger (x')
\end{array} .
\right.
\label{form_factor_momentum}
\ee
Thus, the inverse instanton size plays the role of an ultraviolet cutoff 
of the effective theory. It is important to note that the
way in which the ultraviolet regularization of the effective low--energy 
theory is implemented, namely through form factors restricting the 
virtuality (Euclidean momentum) of the quark fields, is unambiguously
determined by the instanton vacuum. 
\par
The effective action, Eq.(\ref{S_eff}), has been derived
in the large--$N_c$ limit, and can be used to compute
hadronic correlation functions in a $1/N_c$--expansion 
(saddle point approximation) \cite{DP86_prep}. 
Correlation functions of mesonic currents computed in this approximation 
generally show good agreement with phenomenology \cite{DP86,DP86_prep}. 
The nucleon correlation function at large $N_c$ is described by a
saddle point with a non-trivial classical pion field 
(``soliton'') \cite{DPP88}. In the nucleon rest frame it is static and
of ``hedgehog'' form,
\be
U_{\rm class} ({\bf x}) &=& 
\exp\left[ i \frac{x^a \tau^a}{r} P(r)\right],
\hspace{1.5cm} r \; \equiv \; |{\bf x}|, 
\label{hedge}
\ee
where $P(r)$ is called the profile function, with 
$P(r) \rightarrow 0$ for $r \rightarrow \infty$. Integration over
translational and (iso--) rotational zero modes of the saddle point
field gives rise to nucleon states of definite spin/isospin quantum
numbers \cite{DPP88}, with the delta resonance appearing as a 
rotational excitation. This picture of the nucleon gives a good
description of practically all hadronic observables of the octet
and decuplet baryons (for a review see Ref.\cite{Review}). It also
describes well the twist--2 parton distributions of the 
nucleon \cite{DPPPW96,PPGWW98}.
\subsection{Effective operators for QCD twist--4 operators}
\label{subsec_effective_operators}
The instanton vacuum is a variational approximation to the
full QCD partition function and thus allows to evaluate hadronic 
matrix elements of QCD operators, including such involving
the gluon field. It is understood that the QCD 
operators are normalized at the scale of the inverse instanton size, 
$\mu \sim \bar\rho^{-1}$, the scale at which the strong coupling 
constant is fixed when determining the properties of the instanton 
vacuum. 
\par
A convenient method for computing hadronic matrix elements
of QCD operators, in the same spirit as the effective low--energy
theory, Eq.(\ref{S_eff}), is the method of effective operators 
\cite{DPW96}.
Using the same approximations as those which went into the derivation of
the low--energy theory --- the diluteness of the instanton medium,
the zero mode approximation, and the large--$N_c$ limit --- 
it is possible to ``translate'' QCD composite operators into 
effective operators, whose hadronic matrix elements can be computed within 
the effective theory. This method has been used to calculate
various vacuum condensates \cite{PW96} as well as spin--dependent 
and --independent nucleon matrix elements of higher--twist operators
\cite{BPW97}. It was shown in Ref.\cite{BPW97} that basic relations 
between matrix elements of higher--twist operators following 
from the QCD equations of motion are preserved at the level of effective
operators. We now compute in this approach the twist--4 operators 
arising in the $1/Q^2$--corrections to $F_L$ and $F_2$, 
Eqs.(\ref{A_def}), (\ref{B_def}) and (\ref{C_def}).
\par
For a QCD composite operator made purely from quark fields the effective 
operator in leading order of the packing fraction,
$\bar\rho / \bar R$, is simply given by 
the QCD operator with the fields replaced by the massive quark
field of the effective low--energy theory \cite{DPW96}. For a gluon
or quark--gluon operator, one must in addition (after passing to the
Euclidean theory) replace the gluon
fields in the operator by the field of one $I$ ($\bar I$)
and integrate over its collective coordinates; multi--instanton
contributions are suppressed by additional powers of the 
packing fraction. One immediately sees that at this
level the effective operator for the QCD operator $A_{\alpha\beta}$, 
Eq.(\ref{A_def}), is identically zero, since the instanton field is a 
solution to the Euclidean Yang--Mills 
equations,
\be
D_\gamma F_{\gamma\alpha} (x)_{I(\bar I)} &\equiv& 0 .
\label{DF_instanton}
\ee
Thus, the operator $A_{\alpha\beta}$ requires at least two--instanton
contributions to be non-zero, and its matrix element is of order 
$(\bar\rho / \bar R)^4$, {\it i.e.}, parametrically suppressed.
In contrast, the function of the gauge field appearing in the
operator $B_{\alpha\beta}$, Eq.(\ref{B_def}), is non-zero at 
one--instanton level (see below), and its
matrix elements can be of order unity in the packing fraction.
\par
One may ask what would have happened if we had started instead 
from the four--fermionic version of the QCD operator, $A_{\alpha\beta}$, 
Eq.(\ref{A_fourfermion}), which is equivalent to the original
quark--gluon operator by the QCD equations of motion. In 
Section \ref{subsec_fourfermion} we shall explicitly compute the
nucleon matrix element of the four--fermionic operator 
Eq.(\ref{A_fourfermion}) and see that it, too, is suppressed by a 
power of the instanton packing fraction: the matrix element
is of order $M^2$, while that of the operator $B_{\alpha\beta}$
is of order $\bar\rho^{-2}$. Either way, we arrive at the same conclusion: 
In the instanton vacuum the quark--gluon operator proportional to the 
covariant divergence of the gauge field, $A_{\alpha\beta}$, 
Eq.(\ref{A_def}), is suppressed relative to the operator $B_{\alpha\beta}$. 
\par
Also in Section \ref{subsec_fourfermion} we shall investigate the
matrix elements of the four--fermionic operator, $C_{\alpha\beta}$,
Eq.(\ref{C_def}), which appears in the power corrections to $F_2$ only.
Its matrix elements also turn out to be suppressed relative to 
those of the quark--gluon operator, $B_{\alpha\beta}$. 
\par
\begin{figure}
\setlength{\epsfxsize}{15cm}
\setlength{\epsfysize}{13cm}
\epsffile{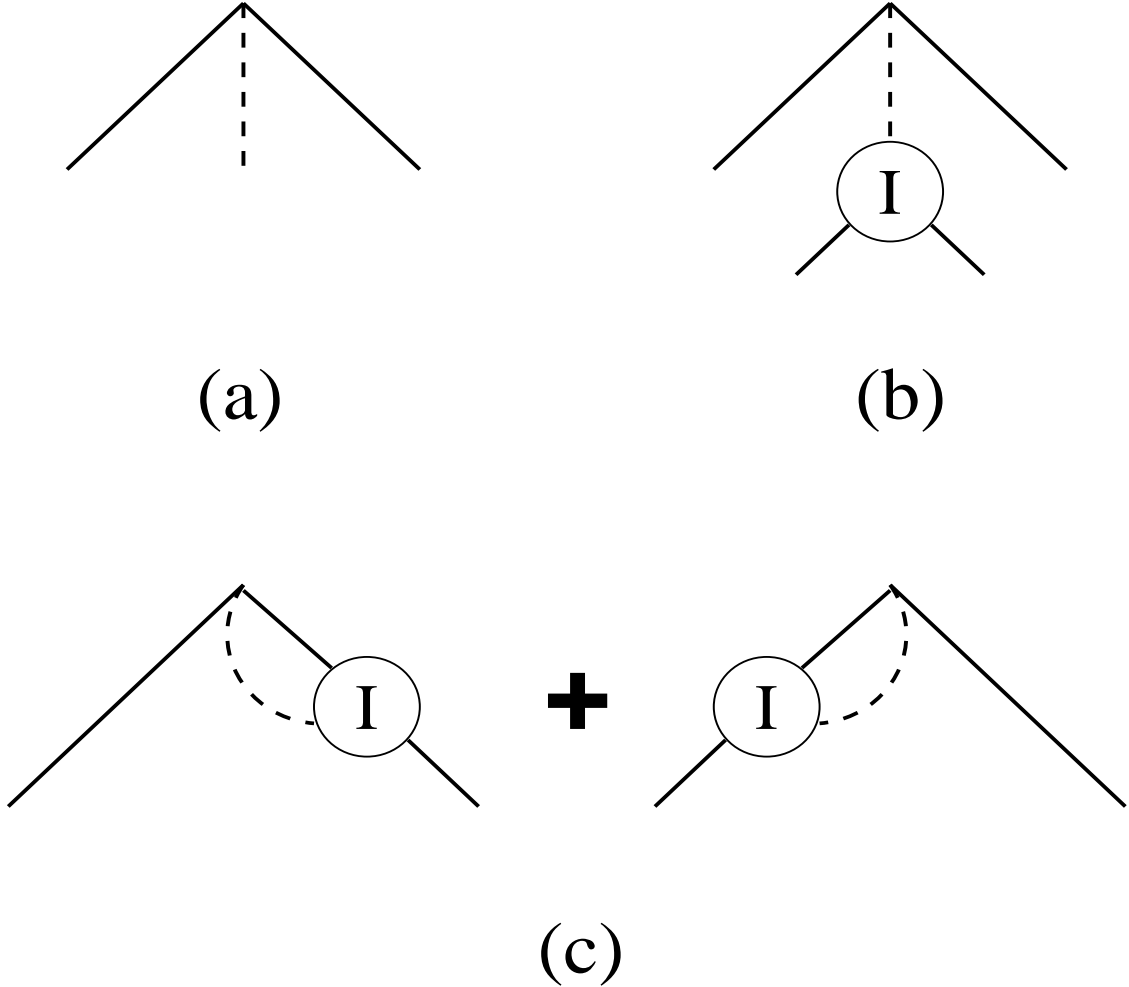}
\caption[]{Graphical illustration of the effective operator approach.  
(a) The QCD quark--gluon operator $B_{\alpha\beta}$, Eq.(\ref{B1_E}),
normalized at $\mu = \bar\rho^{-1} = 600\,{\rm MeV}$. (b) The effective
four--fermion operator, Eq.(\ref{B_fourfermion}). The gluon field has been
replaced by the field of an (anti--) instanton, which couples to quarks
through the zero modes. (c) The two contractions of the four--fermionic
operator giving rise to parametrically large contributions. The loop
integrals here are ``quadratically divergent'' ($\sim \bar\rho^{-2}$). The
sum of the two contractions defines the effective one--body operator,
Eq.(\ref{onebody}).}
\label{fig_ops}
\end{figure}

We now compute the effective operator for $B_{\alpha\beta}$, 
Eq.(\ref{B_def}), in the instanton vacuum, following 
the steps described in Refs.\cite{DPW96,BPW97}.
For use with the Euclidean theory we define Euclidean vector
components of the Minkowskian operator, Eq.(\ref{B_def}), according 
to ($i, j = 1\ldots 3$)
\be
(B_{ij})_E &=& (B_{ij})_M ,
\hspace{1cm}
(B_{44})_E \; = \; -(B_{00})_M ,
\hspace{1cm}
(B_{i4})_E \; = \; -i(B_{i0})_M ,
\ee
and express them in terms of the Euclidean fields and gamma matrices, 
using the conventions given in Appendix \ref{app_eucl}. 
(We suppress the labels denoting Euclidean components in the 
following; unless stated otherwise all formulas in this section
are for Euclidean objects.) It will be convenient in the 
following to separate the parts originating from
the ordinary derivative and the gauge potential in the covariant 
derivative in Eq.(\ref{B_def}):
\be
B_{\alpha\beta, \; f} (x) &=& 
\left[ -i\psi^\dagger_f (x) \frac{\lambda^a}{2}
\gamma_\gamma \gamma_5 \, i\! \partialboth_\alpha
\psi_f (x) \right] \Fdual_{\beta\gamma} (x)
\label{B1_E}
\\
&& + \; (-i)\psi^\dagger_f (x) \gamma_\gamma \gamma_5 
\left\{ \frac{\lambda^b}{2} , \frac{\lambda^c}{2} \right\}_+ \psi_f (x) 
\; A_{\alpha}^b (x) \Fdual_{\beta\gamma}^c (x) 
\label{B2_E} 
\ee
(symmetrization and trace subtraction will not be explicitly shown in the
following). This separation refers explicitly to the ``singular'' gauge of
the instanton field.  In the first part, Eq.(\ref{B1_E}), we have performed
an integration by parts 
($\partialboth \equiv \partialright - \partialleft$), dropping total
derivatives which do not contribute to the forward matrix
element. Following Refs.\cite{DPW96,BPW97} we have to substitute in
Eqs.(\ref{B1_E}) and (\ref{B2_E}) the gauge field of an $I$ ($\bar I$) and
integrate over its collective coordinates. For the part Eq.(\ref{B1_E}) we
need the dual field strength, which for $I$ ($\bar I$) in standard
orientation and centered at zero takes the form
\be
\Fdual^a_{\beta\gamma} (x)_{I (\bar I)} 
&=& \pm\, F^a_{\beta\gamma} (x)_{I (\bar I)} 
\;\; = \;\; \pm \, (\eta^\mp )^a_{\mu\nu} 
f_{\mu\nu, \beta\gamma} (x) ,
\nonumber \\
f_{\mu\nu, \beta\gamma} (x)
&=& \frac{8  \rho^2 }{(x^2 + \rho^2 )^2}
\left( \frac{x_\mu x_\beta}{x^2} \delta_{\gamma\nu} 
+ \frac{x_\nu x_\gamma}{x^2} \delta_{\mu\beta} 
- \half \delta_{\mu\beta} \delta_{\gamma\nu} \right), 
\label{Fdual_inst}
\ee
where $(\eta^\mp )^a_{\mu\nu} \equiv \bar\eta^a_{\mu\nu}, \eta^a_{\mu\nu}$
are the 't Hooft symbols. For operators of this type the form of the
effective operator has been derived in Ref.\cite{BPW97}.  After integration
over instanton coordinates one obtains from the part Eq.(\ref{B1_E}) the
following Euclidean effective operator:
\be
\lefteqn{
\effop{B}_{\alpha\beta, \; f} (x)
\;\; = \;\;  \left[ -i \psi^\dagger_f (x) \frac{\lambda^a}{2}
\gamma_\gamma \gamma_5 \, i \!\partialboth_\alpha \, \psi_f (x) \right] } && 
\nonumber \\
&& \times \frac{i M}{N_c} \; \int d^4 z \; \sum_{g, g'}^{N_f}
f_{\mu\nu, \beta\gamma} (x - z)
\left[ \psi^\dagger_g (z) \, F(-\!\partialleftsq ) \, \frac{\lambda^a}{2}
\sigma_{\mu\nu} \gamma_5 \, U^{\gamma_5}_{gg'} (z) \, 
F(-\!\partialrightsq ) \, \psi_{g'} (z) \right] .
\hspace{1cm}
\label{B_fourfermion}
\ee
Here $\sigma_{\mu\nu} \equiv (i/2) [\gamma_\mu , \gamma_\nu]$.  The form of
the effective operator is intuitively plausible: The gluon field in the QCD
operator, Eq.(\ref{B1_E}), has been replaced by the field of a single 
$I ({\bar I})$, which couples to the fermions through the zero modes. This
induced fermion vertex is chirally odd ($\propto \sigma_{\mu\nu}$) and
accompanied by a factor $M$; however, chiral invariance is maintained due
to the presence of the pion field.\footnote{The vertex involving the pion
field in Eq.(\ref{B_fourfermion}) is obtained from the original 
$N_f \times N_f$--fermionic instanton--induced vertex by making use of the
saddle--point condition of the effective theory at large $N_c$; see
\cite{BPW97} for details.}  The presence of $\gamma_5$ in the
instanton--induced vertex in Eq.(\ref{B_fourfermion}) is a consequence of
adding instanton [$\propto (1 + \gamma_5)/2$] and antiinstanton 
[$\propto (1 - \gamma_5)/2$] contribution with different sign, {\it cf.}\
Eq.(\ref{Fdual_inst}).  The effective operator can graphically be
represented as in Fig.\ref{fig_ops} (b).
\par
In a similar way one can derive the contribution of the part
Eq.(\ref{B2_E}) to the effective operator. The anticommutator of Gell--Mann
matrices contains a color--singlet as well as a color--octet piece. We
shall not write down these contributions here. It turns out that the
``derivative'' part, Eq.(\ref{B1_E}), gives the (by far) numerically
dominant contribution to the matrix element, so we shall drop the
contributions from Eq.(\ref{B2_E}).
\par
The effective operator, Eq.(\ref{B_fourfermion}), can now be inserted in
correlation functions of hadronic currents computed within the effective
low--energy theory. We are interested only in the leading contributions in
$\bar\rho / \bar R$, {\it viz.} $M \bar\rho$ to the hadronic matrix
element. For dimensional reasons, the instanton--induced vertex in
Eq.(\ref{B_fourfermion}) is proportional to a positive power of the
instanton size, $\bar\rho^2$. A parametrically large contribution to
correlation functions can thus only come from diagrams in the effective
theory containing ``quadratically divergent'' loop integrals, {\it i.e.},
integrals giving rise to a factor of $\bar\rho^{-2}$, which cancels the
factor $\bar\rho^2$ incurred from the instanton--induced vertex (remember
that $\bar\rho$ also plays the role of ultraviolet cutoff of the effective
theory). One can easily see that such quadratic divergences can only arise
from diagrams of the type shown in Fig.\ref{fig_ops} (c), where the quark
propagator in the loop can be taken as the free quark propagator,
$(-i\gamma\cdot\partial )^{-1}$; taking into account the coupling of the
quarks in the loop to the pion field would lead to contributions of higher
order in $M\bar\rho$. Thus we may replace the four--fermionic operator
Eq.(\ref{B_fourfermion}) by a one--body operator representing the sum of
the two contractions of Fig.\ref{fig_ops} (c).  Due to the ``separable''
form of the instanton--induced vertex it is of the form:
\be
\effop{B}_{\alpha\beta, \; f} (x)_{\rm one-body}
&=& \frac{i M}{2} \; \sum_{f'}^{N_f} \left[ \psi^{\dagger}_f (x)  
\,\, \Gamma^{(1)}_{\alpha\beta} (i\!\partialleft ) \,\,
U^{\gamma_5}_{ff'} (x) \, F(-\!\partialrightsq ) \, \psi_{f'} (x)
\right.
\nonumber \\
&& \left. \;\;\;\;\;\;\;\; + \;\; 
\psi^{\dagger}_{f'} (x) \, F(-\!\partialleftsq )
\, U^{\gamma_5}_{f'f} (x) 
\,\, \Gamma^{(2)}_{\alpha\beta} (-i\!\partialright ) \,\,
\psi_f (x) 
\right] .
\label{onebody}
\ee
The vertices $\Gamma^{(1)}_{\alpha\beta}$ and $\Gamma^{(2)}_{\alpha\beta}$
can be determined by computing the loops in the diagrams of
Fig.\ref{fig_ops} (c) in momentum representation:
\be
\left.
\begin{array}{l}
\Gamma^{(1)}_{\alpha\beta} (p)
\\[2ex]
\Gamma^{(2)}_{\alpha\beta} (p)
\end{array}
\right\}
&=& (-i) \fourint{k} 
f_{\mu\nu, \beta\gamma} (k)
\nonumber \\
&& \times
\left\{
\begin{array}{l}
\frac{\displaystyle F[(p - k)^2 ]}{\displaystyle (p - k)^2} (-2 p + k)_\alpha
(p - k)_\delta \left[ \gamma_\gamma \gamma_5 \gamma_\delta 
\sigma_{\mu\nu} \gamma_5 \right]
\\[2ex]
\frac{\displaystyle F[(p + k)^2 ]}{\displaystyle (p + k)^2} (-2 p - k)_\alpha
(p + k)_\delta \left[ \sigma_{\mu\nu} \gamma_5 \gamma_\delta 
\gamma_\gamma \gamma_5 \right] ,
\end{array}
\right.
\label{C_pm_1_2}
\ee
where 
\be
f_{\mu\nu , \beta\gamma} (k)
&\equiv& \int d^4 x \, e^{-ik\cdot x} \;
f_{\mu\nu , \beta\gamma} (x)
\nonumber \\
&=& \bar\rho^2 {\cal G} (k^2 ) 
\left( \frac{k_\mu k_\beta}{k^2} \delta_{\gamma\nu} 
+ \frac{k_\nu k_\gamma}{k^2} \delta_{\mu\beta} 
- \half \delta_{\mu\beta} \delta_{\gamma\nu} \right) ,
\label{F_inst_four}
\ee
\be
{\cal G} (k^2 ) &=& 32 \pi^2 \left[ \left(\frac{1}{2} + \frac{4}{t^2} \right) 
K_0 (t) + \left( \frac{2}{t} + \frac{8}{t^3} \right) K_1 (t) 
- \frac{8}{t^4} \right], \hspace{1cm} t \; \equiv \; \bar\rho\sqrt{k^2} .
\label{G} 
\ee
Since the instanton--induced vertex in Eq.(\ref{B_fourfermion}) is 
$\propto \sigma_{\mu\nu}$, the vertex in the one--body operator,
Eq.(\ref{onebody}), is also chirally odd and can be reduced to Dirac
structures $1$ and $\sigma_{\kappa\lambda}$. Note that, again, chiral
invariance is preserved by the presence of the pion field in
Eq.(\ref{onebody}).  Computing the loop integrals Eq.(\ref{C_pm_1_2}) we
find that both structures occur with the same form factors. The resulting
one--body operator is given by
\be
\lefteqn{ \effop{B}_{\alpha\beta, \; f} (x)_{\rm one-body} }
&& \nonumber \\
&=& (-i)\frac{M}{2} \; \sum_{f'}^{N_f} \left[ 
\psi^\dagger_f (x) \;
( \partialleft_\alpha \partialleft_\beta 
- i \partialleft_\alpha \partialleft_\kappa \sigma_{\kappa\beta} ) \;
{\cal B} (-\!\partialleftsq ) \, U^{\gamma_5}_{ff'}(x) \, 
F(-\!\partialrightsq )
\; \psi_{f'} (x)
\right.
\nonumber \\
&& \;\; + \; \left. \psi^\dagger_{f'} (x) \;
F(-\!\partialleftsq ) \, U^{\gamma_5}_{f'f}(x) \, 
{\cal B}(-\!\partialrightsq ) 
\; ( \partialright_\alpha \partialright_\beta
+ i \partialright_\alpha \partialright_\kappa 
\sigma_{\kappa\beta} ) \; \psi_f (x)
\right] ,
\label{B_onebody}
\ee
where ${\cal B}(p^2 )$ is a dimensionless scalar form factor defined as
\be
{\cal B}(p^2 ) &=& \bar\rho^2 \fourint{k} 
{\cal G} (k^2 ) \frac{F[(p - k)^2 ]}{(p - k)^2} 
\nonumber \\
&& \times \left[
-\frac{19}{12} \frac{k}{p} C^{(1)}_1 (\cos\theta )
+ \left( \frac{1}{2} \frac{k^2}{p^2} + 1 \right) C^{(1)}_2 (\cos\theta ) 
- \frac{1}{3} \frac{k}{p} C^{(1)}_3 (\cos\theta )
\right] .
\;\;\;\;\;\;
\label{C_integral}
\ee
Here the $C^{(1)}_n$ denote the 4--dimensional spherical harmonics
(Gegenbauer polynomial of index $1$):
\be
C^{(1)}_n (\cos\theta ) &=& \frac{\sin (n + 1) \theta}{\sin\theta} ,
\hspace{1cm}
\cos\theta \;\; = \;\; \frac{k\cdot p}{|k| |p|} .
\ee
The integral in Eq.(\ref{C_integral}) is of order $\bar\rho^{-2}$
(``quadratically divergent''), as a result of which ${\cal B}(p)$ is of
order unity in $M\bar\rho \sim (\bar\rho / \bar R )^2$.  Numerical
evaluation of the integral gives
\be
{\cal B}(p^2 = 0) &=& 0.78 ,
\label{G_at_0}
\ee
and shows that ${\cal B}(p^2 ) \sim (\bar\rho^2 p^2)^{-5/2}$ for large
$p^2$.
\par
To summarize, the parametrically large contributions to the matrix element
of the quark--gluon operator $B_{\alpha\beta}$, Eq.(\ref{B_def}), are
contained in the ``quadratically divergent'' part of the contractions
Fig.\ref{fig_ops} (c), which can be represented by a chirally odd
one--body quark operator in the background of the pion field.
\par
The effective one--body operator, Eq.(\ref{B_onebody}), can also be written
in a different form. Making use of the equations of motion of the effective
low--energy theory, {\it cf.}\ Eq.(\ref{S_eff}),
\be
-i M \; \sum_{f'}^{N_f} \psi^\dagger_f \; \ldots \;
F(-\partialleftsq ) U^{\gamma_5}_{ff'} F(-\partialrightsq ) 
\; \psi_{f'} &=& 
\psi^\dagger_f \; \ldots \; i \gamma_\mu \partialright_\mu
\; \psi_f , 
\nonumber \\
-i M \; \sum_{f'}^{N_f} \psi^\dagger_{f'} \; F(-\partialleftsq ) 
U^{\gamma_5}_{f'f} F(-\partialrightsq ) \;
\ldots \; \psi_f &=& 
\psi^\dagger_f \; (-i) \gamma_\mu\partialleft_\mu \; \ldots \; 
\psi_f 
\ee
[the ellipses denote additional functions of $\partialright$ or
$\partialleft$, respectively], we can eliminate the pion field in
Eq.(\ref{B_onebody}) and write the operator in a manifestly chirally
invariant form. Integrating by parts and simplifying the product of gamma
matrices, we finally obtain an operator
\be
\effop{B}_{\alpha\beta, \; f}
&=& \psi^\dagger_f \; F^{-1} (-\partial^2 ) \, {\cal B}(-\partial^2 ) 
\; \partial^2 \; i\partial_{\left\{ \alpha \right.} 
\gamma_{\left.\beta\right\} } \, \psi_f 
\; - \; \mbox{trace} .
\label{B_simple}
\ee
The meaning of the ``inverse'' form factor here is clear in momentum
representation, {\it cf.}\ Eq.(\ref{form_factor_momentum}). In this
representation of the effective operator the effects of the gluon field of
the instanton are contained in additional contracted derivatives acting on
the quark fields relative to the spin--2 twist--2 operator,
Eq.(\ref{twist2_def}). The operator Eq.(\ref{B_simple}) measures the
``average virtuality'' of the quark in the nucleon.  This is what one
should expect on general grounds in QCD: By the QCD equations of motion,
the explicit gauge fields in twist--4 operators in the ``collinear basis''
can be traded for transverse derivatives of the quark fields, which in turn
are related to the virtuality of the quark in the nucleon \cite{EFP82}.
\par
For momenta of the quark fields $p^2 \ll \bar\rho^{-2}$ we can neglect the
momentum dependence of the form factors in Eq.(\ref{B_simple}), so that the
effective operator reduces to a local operator:
\be
\effop{B}_{\alpha\beta, \; f}
&\approx& {\cal B}(0) \; 
\psi^\dagger_f \, \partial^2 \; i\partial_{\left\{ \alpha \right.} 
\gamma_{\left.\beta\right\} } \, \psi_f 
\; - \; \mbox{trace} .
\label{B_local}
\ee
Upon returning to the Minkowskian theory (see Appendix \ref{app_eucl}) this
operator becomes
\be
\effop{B}_{\alpha\beta, \; f} |_{\rm Minkowski}
&=& {\cal B}(0) \; \bar\psi_f \; \partial^2 
\; i\partial_{\left\{ \alpha \right.} 
\gamma_{\left.\beta\right\} } \, \psi_f 
\; - \; \mbox{trace} .
\label{B_Minkowski}
\ee
It is interesting to note that the reduced matrix element of this operator
in a ``constituent'' quark state [{\it cf.}\ Eq.(\ref{me_A_B})], {\it
i.e.}, an on--shell quark state with $(p^2)_{\rm Minkowski} = M^2$, has a
negative value (for any flavor $f$):
\be
B_{\rm quark} &=& -M^2 {\cal B}(0) \;\; < 0 .
\ee
However, as we shall see below, in the nucleon the flavor--singlet matrix
element is positive, since the quarks in the nucleon have on average
spacelike momenta, so $(p^2)_{\rm Minkowski}$ effectively becomes negative.
This exercise shows that the effective operator, Eq.(\ref{B_simple}),
should not be used out of context: It refers explicitly to the effective
low--energy theory derived from the instanton vacuum, where the quarks
inside hadrons have virtualities (Euclidean momenta) up to the ultraviolet
cutoff, $\bar\rho^{-1}$. In particular, the notion of ``average
virtuality'' of the quarks has no absolute meaning but is defined by the
ultraviolet regularization of the effective theory.
\par
We remark that for the contribution from the part Eq.(\ref{B2_E}) to the
effective operator one can also derive an effective one--body operator
similar to Eq.(\ref{B_onebody}). Explicit calculation shows that for this
part the form factors are numerically an order of magnitude smaller than in
Eq.(\ref{B_onebody}), so we can drop this contribution.
\subsection{Nucleon matrix element of the quark--gluon operator $B$}
\label{subsec_nucleon}
We now compute the nucleon matrix element of the operator
$B_{\alpha\beta}$, Eq.(\ref{me_A_B}), in the effective chiral theory.  To
obtain an explicit expression for the reduced matrix element, $B_f$, we
contract both sides of Eq.(\ref{me_A_B}) with a Minkowskian light--like
vector, $n^\mu$ ($n^2 = 0, \; n^\mu p_\mu = M_N )$:
\be
B_f &=& \frac{1}{2 \, M_N^2}
n^\alpha n^\beta \; \langle p | B_{\alpha\beta , f} | p \rangle .
\label{me_A_B_projected}
\ee
This expression is manifestly covariant and can therefore be evaluated in
any frame. In our large--$N_c$ approach it is convenient to work in the
nucleon rest frame, where the nucleon is described by a static classical
pion of the form Eq.(\ref{hedge}), subject only to adiabatic rotations and
translations whose quantization gives rise to the nucleon quantum numbers.
We are interested in the parametrically leading contribution to the matrix
element in the instanton packing fraction, $\bar\rho / \bar R$, which can
be computed passing to the Euclidean theory and replacing the operator
$B_{\alpha\beta}$ by the effective operator obtained from instantons,
Eq.(\ref{B_simple}).
\par
A well--developed technique exists for computing matrix elements of quark
operators within the $1/N_c$--expansion. It can directly be applied to
matrix elements of the instanton--induced effective operators derived from
twist--4 QCD operators, see Ref.\cite{BPW97} for details.  For simplicity
we consider the effective low--energy theory with two light quark flavors,
$N_f = 2$. The isosinglet part of the matrix element
Eq.(\ref{me_A_B_projected}) appears in leading order of the
$1/N_c$--expansion, the isovector part only in next--to--leading order,
{\it i.e.}, after expanding to first order in the angular velocity of the
soliton. Noting that the form factor ${\cal B}$ in the one--body operator
is of order unity in $N_c$, standard $N_c$ counting tells us that
\be
B_S &\equiv& B_u^{\rm p} + B_d^{\rm p} 
\;\; = \;\;\;\;\;\; B_u^{\rm n} + B_d^{\rm n}  \;\;\; 
\sim \;\; 1, 
\nonumber \\
B_{NS} &\equiv& B_u^{\rm p} - B_d^{\rm p} \;\; = \;\; -(B_u^{\rm n} - 
B_d^{\rm n} )  
\;\; \sim \;\; 1/N_c , 
\ee
(the isoscalar matrix element is of the same order in $N_c$ as the momentum
fraction of quarks, which is order unity). Incidentally, this means that
the target mass corrections, Eqs.(\ref{M_L_ex_target}) and
(\ref{M_2_target}), are of the same order in $1/N_c$ as the dynamical
twist--4 contributions, since $M_N \sim N_c$ and $L^{(4)}_S \sim 1/N_c^2$,
$L^{(4)}_{NS} \sim 1/N_c^3$, as follows from the analysis of twist--2
parton distributions in the large--$N_c$ limit in Ref.\cite{DPPPW96}. Thus,
it makes sense to consider both kinds of corrections simultaneously in the
large--$N_c$ limit.
\par
We now compute the isoscalar part, $B_S$, which is leading in the
$1/N_c$--expansion.  Since $M\bar\rho \propto (\bar\rho / \bar R)^2$, it is
sufficient to consider only the leading contribution in $M\bar\rho$ --- the
leading ``ultraviolet divergence'', which is contained in the lowest--order
term in the expansion of the matrix element in gradients of the classical
pion field of the nucleon (see Ref.\cite{DPPPW96} for details).  To derive
the gradient expansion we express the nucleon matrix element of
Eq.(\ref{B_simple}) in terms of the Euclidean Green function of quarks in
the background pion field, {\it cf.}\ Eq.(\ref{S_eff}),
\be
G({\bf x}, x_4; {\bf y}, y_4) &=& \langle {\bf x}, x_4 |
\frac{1}{- i\gamma\cdot\partial 
- i M F(-\partial^2 ) U^{\gamma_5}_{\rm class} F(-\partial^2 )}
| {\bf y}, y_4 \rangle ,
\label{G_background}
\ee
where $|{\bf x}, x_4 \rangle$ denotes a set of formal ``position
eigenstates''. After integration over the soliton center in order to
project on the nucleon state with zero three--momentum, and over soliton
orientations in isospin space, one has (Euclidean conventions)
\be
B_S &=& -\frac{N_c}{M_N} \; n_\alpha n_\beta \; \int d^3 x
\; {\rm tr}\left[ \; \langle {\bf x}, 0| \,
 F^{-1}(-\partial^2 ) \, {\cal B}(-\partial^2 ) \,
\partial^2 \, i\partial_\alpha \gamma_\beta
\right. \nonumber \\
&& \left. \times \frac{1}{- i\gamma\cdot\partial 
- i M F(-\partial^2 ) U^{\gamma_5}_{\rm class} F(-\partial^2 )}
\, | {\bf x}, 0 \rangle \; \right] , \;\;\;\;
\label{trace}
\ee
where $n_\alpha$ are the Euclidean components of the light--like vector,
and the trace runs over Dirac spinor and flavor indices.  Expanding in
gradients of the classical pion field, and evaluating the R.H.S.\ of
Eq.(\ref{trace}) using plane--wave states, we find in leading order 
($i, j = 1, \ldots 3$):
\be
B_S &\approx& \frac{2 N_c M^2 I_2}{M_N} \; n_i n_j \,\int d^3 x \; 
{\rm tr} [\partial_i U^\dagger_{\rm class} ({\bf x}) 
\partial_j U_{\rm class} ({\bf x})] 
\nonumber \\
&=& \frac{2 N_c M^2 I_2}{3 M_N} \; \int d^3 x \; 
{\rm tr} [\partial_i U^\dagger_{\rm class} ({\bf x}) 
\partial_i U_{\rm class} ({\bf x})] .
\label{B_gradient}
\ee
In the last step we have averaged over the arbitrary orientation of the
three--dimensional vector, 
${\bf n}$: $n_i n_j \rightarrow \delta_{ij}/3$. Here, $I_2$ denotes a quark
loop integral of dimension $(\mbox{mass})^2$:
\be
I_2 &=& \fourint{p} \frac{{\cal B}(p^2 ) F(p^2 )}{p^2}
\nonumber \\
&& \times \left\{ F^2 (p^2 )  - 2 p^2 F (p^2 ) F'(p^2 ) + \frac{2}{3} p^4
\left[ F^{'2}(p^2 ) + F(p) F^{''}(p^2 ) \right] \right\} ,
\label{I}
\ee
where $p$ is Euclidean ($p^2 > 0$) and $F' \equiv dF/d(p^2)$.  This
integral is a measure of the ``average virtuality'' of the quarks and
antiquarks in the nucleon, as determined by the instanton--induced form
factors, $F(p)$. Parametrically, $I_2 \sim \bar\rho^{-2}$, {\it i.e.}, the
average virtuality is of the order of the square of the ultraviolet cutoff,
$\bar\rho^{-1}$. Note that in the expression Eq.(\ref{I}) we have set $M =
0$ and kept only the leading contribution in $M\bar\rho$.
\par
The result for the gradient expansion of $B_S$, Eq.(\ref{B_gradient}),
should be compared to that the matrix element of the twist--2 spin--2
operator, Eq.(\ref{twist2_def}):
\be
L^{(2)}_S &=& \frac{2 N_c M^2 I_0}{3 M_N} \; \int d^3 x \; 
{\rm tr} [\partial_i U^\dagger_{\rm class} ({\bf x}) 
\partial_i U_{\rm class} ({\bf x})] ,
\label{L_2_gradient}
\ee
where $I_0$ is a loop integral, which is parametrically of order 
$\log (M\bar\rho)$:
\be
I_0 &=& \fourint{p} 
\frac{F^2 (p^2 ) \left\{
F^2 (p^2 )  - 2 p^2 F (p^2 ) F'(p^2 ) + \frac{2}{3} p^4
 [ F^{'2}(p^2 ) + F(p) F^{''}(p^2 )] \right\} }
{[p^2 + M^2 F^4 (p^2)]^2} . \;\;\;\;
\label{I_0}
\ee
Up to terms proportional to derivatives of the form factors it can be
identified with the pion decay constant in the effective chiral theory
\cite{DP86}:
\be
F_\pi^2 &\approx& 4 M^2 N_c \, I_0 
\ee
($F_\pi^2 = 93\, {\rm MeV}$). Comparing Eq.(\ref{B_gradient}) with 
Eq.(\ref{L_2_gradient}) we obtain
\be
\frac{B_S}{L^{(2)}_S} &=& \frac{I_2}{I_0} \;\; = \;\;
\frac{4 N_c M^2 I_2}{F_\pi^2} .
\label{B_from_A2}
\ee 
Numerically we find $I_2 = 0.0053\, \bar\rho^{-2}$. Using the standard
parameters for the average instanton size and the dynamical quark mass,
$\bar\rho^{-1} = 600\,{\rm MeV}$ and $M = 350\,{\rm MeV}$ \cite{DP86}, and
setting $L^{(2)}$ equal to unity, which is the result obtained in the
instanton vacuum in leading order of $\bar\rho / \bar R$ \cite{DPPPW96}, we
obtain from Eq.(\ref{B_from_A2}) a value
\be
B_S &=& 0.9 \, \bar\rho^{-2} \;\; = \;\; (570 \, {\rm MeV})^2 .
\label{B_final}
\ee
Thus, $B_S$ has a large numerical value due to the smallness of the
instanton size. In particular, $B_S$ is of order unity in the packing
fraction of the instanton medium,
\be
B_S &\sim& \bar\rho^{-2} \;\; \sim \;\; 
\left(\frac{\bar\rho}{\bar R}\right)^0 .
\ee
\par
A comment is in order here concerning the accuracy of the gradient
expansion of the nucleon matrix elements, which takes into account the
leading contributions in $M\bar\rho$.  It is known that in the case of the
twist--2 spin--2 operator the gradient expansion, Eq.(\ref{L_2_gradient}),
underestimates the result of the exact calculation of the matrix element,
$L^{(2)}_S = 1$, by a factor of $2/3$ (see Ref.\cite{DPPPW96} for a
detailed discussion). This is apparent when one compares
Eq.(\ref{L_2_gradient}) with the well-known expression for the nucleon mass
in gradient expansion
\be
M_N &=& \frac{F_\pi^2}{4} \; \int d^3 x \; 
{\rm tr} [\partial_i U^\dagger_{\rm class} ({\bf x}) 
\partial_i U_{\rm class} ({\bf x})] .
\label{M_N_gradient}
\ee
The reason for this ``2/3--paradox'' is that the momentum sum rule,
$L^{(2)}_S = 1$, requires the equations of motion for the classical pion
field of the nucleon to be satisfied; however, the latter have a
non-trivial solution only if finite terms in $M\bar\rho$ are taken into
account.  An analogous factor of $2/3$ occurs in the gradient expansion
expression for $B_S$, Eq.(\ref{B_gradient}). If we had compared our result
Eq.(\ref{B_gradient}) to the expression for nucleon mass in gradient
expansion, Eq.(\ref{M_N_gradient}), rather than the twist--2 matrix
element, Eq.(\ref{L_2_gradient}), we would have obtained an estimate for
$B_S$ smaller by a factor of $2/3$. We believe the comparison with the
matrix element of the twist--2 spin--2 operator to give the more realistic
numerical estimate. Note that this question can ultimately be decided only
in a theory which correctly takes into account all orders of $M\bar\rho$,
that is, of the instanton packing fraction, $\bar\rho / \bar R$. Since our
effective operator has been derived only in the leading order of 
$\bar\rho / \bar R$, it would not be sensible to ``improve'' the
calculation of its nucleon matrix element beyond this accuracy. At the
present level, the difference between the two estimates should be seen as
indicating the theoretical uncertainty of our result.
\par
In addition, we have estimated the nucleon matrix element $B_S$ by
computing separately the contribution of the bound--state level and the
Dirac continuum of quarks to the nucleon matrix element. The former can be
computed exactly, using the known bound--state level wave function; the
latter can be estimated using an ``interpolation formula'' (see
Ref.\cite{BPW97}), which becomes exact in various limiting cases and
correctly reproduces the leading ultraviolet divergence given by
Eq.(\ref{B_gradient}). The numerical result lies between the two possible
values obtained from gradient expansion, so all estimates are well
consistent.
\subsection{Nucleon matrix elements of the four--fermionic operators}
\label{subsec_fourfermion}
We now compute the nucleon matrix element of the four--fermionic
(``diquark'') twist--4 operator, $C_{\alpha\beta}$, Eq.(\ref{C_def}), which
arises from contributions to the DIS cross section of the type shown in
Fig.\ref{fig_twist4} (b).
\par
First we note that the four--fermionic twist--4 operator,
$C_{\alpha\beta}$, Eq.(\ref{C_def}), being a local operator, does not have
contractions with closed loops which give rise to parametrically large
($\sim \bar\rho^{-2}$) integrals; the corresponding loop integrals are of
type $\int d^4 k\, (k_\mu/k^2 )$ and vanish because of rotational
invariance. This is in contrast to the effective four--fermionic operator
obtained from the quark--gluon operator, $B_{\alpha\beta}$,
Eq.(\ref{B_fourfermion}), which has a ``quadratically divergent''
contraction, see Fig.\ref{fig_ops} (c). Consequently, the four--fermionic
twist--4 operator, $C_{\alpha\beta}$, cannot be represented by a one--body
quark operator of the type Eq.(\ref{B_onebody}) in the effective
theory. Its matrix element in a hadron is not related to the virtuality of
quarks in the bound state; it rather measures two--body correlations
between quarks.  For this reason we expect the nucleon matrix element of
$C_{\alpha\beta}$ to be not of order $\bar\rho^{-2}$, but of order $M^2$,
{\it i.e.}, parametrically small.
\par
We are interested in the isosinglet matrix element of the operator
Eq.(\ref{C_def}), {\it i.e.}, the average of proton and neutron matrix
elements. Because of isospin invariance
\be
\lefteqn{
\frac{1}{2} \sum_{f,f'} e_f e_{f'} \left[ 
  \langle {\rm p} | C_{\alpha\beta , ff'} | {\rm p} \rangle 
+ \langle {\rm n} | C_{\alpha\beta , ff'} | {\rm n} \rangle \right] } &&
\nonumber \\
&=& \frac{1}{4} (e_u + e_d )^2 
\langle {\rm p} | C_{\alpha\beta , SS} | {\rm p} \rangle 
\; + \; \frac{1}{12} (e_u - e_d )^2
\langle {\rm p} | C_{\alpha\beta , VV} | {\rm p} \rangle ,
\label{C_singlet}
\ee
where
\be
C_{\alpha\beta , SS} &=& g^2 \left( \sum_f \bar\psi_f \frac{\lambda^a}{2} 
\gamma_\alpha\gamma_5 \psi_f \right)
\left( \sum_{f'} \bar\psi_{f'} 
\frac{\lambda^a}{2} \gamma_\beta\gamma_5 \psi_{f'} \right) ,
\label{C_SS} 
\\
C_{\alpha\beta , VV} &=& 
g^2 \left( \sum_f \bar\psi_f \tau^b
\frac{\lambda^a}{2} 
\gamma_\alpha\gamma_5 \psi_f \right)
\left( \sum_{f'} \bar\psi_{f'} \tau^b 
\frac{\lambda^a}{2} \gamma_\beta\gamma_5 \psi_{f'} \right)
\label{C_VV} 
\ee
are manifestly flavor--singlet operators (the sum over $a = 1, \ldots 8$
and $b = 1,\ldots 3$ is implied).
\par
Since $g^2 \sim 1/N_c$ one finds that the reduced nucleon matrix elements
[{\it cf.}\ Eq.(\ref{me_A_B})] of the four--fermionic operator, $C_{SS}$
and $C_{VV}$, are of order $N_c^0$, and thus of the same order as the
matrix elements of the twist--4 quark--gluon operators. The nucleon matrix
elements can be computed in the chiral soliton picture of the nucleon using
standard techniques.  After integrating over soliton translations and
rotations, we again express the nucleon matrix elements in terms of the
quark Green function in the background pion field, Eq.(\ref{G_background})
(Euclidean conventions):
\be
\left.
\begin{array}{c} 
C_{SS} \\[1ex]
C_{VV}
\end{array}
\right\} 
&=& -\frac{g^2 N_c^2}{2 M_N} \; n_\alpha n_\beta \; \int d^3 x
\; {\rm tr} [ G({\bf x}, 0; {\bf x}, 0) 
\left\{ \begin{array}{c} 1 \\[1ex] \tau^b \end{array} \right\} 
\gamma_\alpha \gamma_5 G({\bf x}, 0; {\bf x}, 0) 
\left\{ \begin{array}{c} 1 \\[1ex] \tau^b \end{array} \right\} 
\gamma_\beta \gamma_5 ] .
\nonumber \\
\label{fourfermion_trace}
\ee
Expanding in derivatives of the pion field we now find in leading order
\be
\left.
\begin{array}{c} 
C_{SS} \\[1ex]
C_{VV}
\end{array}
\right\} 
&\approx& -\frac{4 g^2 N_c^2 M^4}{3 M_N} 
\left( \fourint{k} \frac{F^4 (k)}{[k^2 + M^2 F^4 (k)]^2} \right)
\left( \fourint{k'} \frac{F^4 (k')}{[k'^2 + M^2 F^4 (k')]^2} \right)
\nonumber \\
&& \times 
\left\{ \begin{array}{c} \! + \! \\[1ex] \! - \! \end{array} \right\} 
\int d^3 x \; 
{\rm tr} [ \partial_i U^\dagger_{\rm class} ({\bf x}) 
\partial_i U_{\rm class} ({\bf x}) ] .
\label{fourfermion_grad}
\ee
The Euclidean momentum integrals here are parametrically of order 
$\log (M\bar\rho )$ (``logarithmically divergent'') and, up to terms
involving derivatives of the form factors $F(k)$, which are not essential
here, can be related to the pion decay constant, see Subsection
\ref{subsec_nucleon}.  Comparing Eq.(\ref{fourfermion_grad}) with the
gradient expansion result for the matrix element of the twist--2 spin--2
operator, $L^{(2)}_S$, Eq.(\ref{L_2_gradient}), we find
\be
C_{SS} &=& -C_{VV} \;\; = \;\; -\frac{g^2 F_\pi^2}{2} \, L_S^{(2)} .
\label{C_from_fpi}
\ee
Thus, parametrically
\be
C_{SS}, \; C_{VV} &\sim& M^2 \log (M\bar\rho)
\;\; \sim \;\; \left(\frac{\bar\rho}{\bar R}\right)^4 .
\ee
Inserting Eq.(\ref{C_from_fpi}) in Eq.(\ref{C_singlet}), taking into
account the quark charges, and using $L^{(2)}_S = 1$, we finally obtain
\be
\frac{1}{2} \sum_{f,f'} e_f e_{f'} \left[ 
  \langle {\rm p} | C_{\alpha\beta , ff'} | {\rm p} \rangle 
+ \langle {\rm n} | C_{\alpha\beta , ff'} | {\rm n} \rangle \right]
&=& \frac{g^2 F_\pi^2}{36} \;\; \approx \;\; (50\, {\rm MeV})^2 .
\ee
In the last step we have taken a value of $g^2$ at the scale $\rho^{-1} =
600\, {\rm MeV}$ of $g^2 \approx 8.8$, corresponding to 
$\beta \equiv 8\pi^2/g^2 = 9.0$ \cite{DPW96}. Thus the matrix element of
the four--fermionic operator, $C_{\alpha\beta}$, is significantly smaller
than that of the quark--gluon operator, $B_{\alpha\beta}$, in spite of the
fact that the coupling constant at the low scale is rather sizable. The
reason for this is the different parametric order of the matrix elements:
$M^2$ for the four--fermionic operator vs.\ $\bar\rho^{-2}$ for the
quark--gluon operator.
\par
In Subsection \ref{subsec_effective_operators} we saw that in the instanton
vacuum the matrix element of the quark--gluon operator $A_{\alpha\beta}$,
Eq.(\ref{A_def}), is suppressed, since the function of the gauge field
appearing there is zero in the field of one instanton, {\it cf.}\
Eq.(\ref{DF_instanton}).  It is instructive to compute the nucleon matrix
element of the four--fermionic operator, Eq.(\ref{A_fourfermion}), which is
obtained from $A_{\alpha\beta}$ by applying the QCD equations of motion;
this allows us to check the consistency of our approximations.  The
calculation is analogous to that for the operator $C_{\alpha\beta}$, with
only the singlet times singlet structure, Eq.(\ref{C_SS}), present. We find
\be
A_{S, \; {\rm four-fermion}} &=& \frac{g^2 F_\pi^2}{2} L^{(2)}_S
\;\; \sim \;\; M^2 \log (M\bar\rho)
\;\; \sim \;\; \left(\frac{\bar\rho}{\bar R}\right)^4 ,
\label{A_S_nucleon}
\ee
which is consistent with the fact that the corresponding quark--gluon
operator is zero at one--instanton level, Eq.(\ref{DF_instanton}). With
$g^2 = 8.8$ (see above) and $L^{(2)}_S = 1$ Eq.(\ref{A_S_nucleon}) gives
$A_{S, \; {\rm four-fermion}} = (195\, {\rm MeV})^2$, which is an order of
magnitude smaller than the corresponding matrix element of the ``true''
quark--gluon operator, $B_S$, Eq.(\ref{B_final}). This shows that the
parametric suppression of $A_S$ relative to $B_S$ is indeed borne out by
the numerical values.
\section{Comparison with DIS data}
\setcounter{equation}{0}
\label{sec_comparison}
We now want to confront our results for the twist--4 matrix elements with
the data on $F_L$ and $F_2$ from deep--inelastic scattering at moderate
$Q^2$. We take the usual ``phenomenological'' attitude towards power
corrections, namely, assume that there exists a range of $Q^2$ (typically
$1 \ldots 10\, {\rm GeV}^2$) sufficiently large so that a perturbative QCD
treatment with individual power corrections is justified, but sufficiently
low for the power corrections to be noticeable. We shall not be concerned
here with theoretical questions related to the principal accuracy of the
perturbation series, renormalon ambiguities {\it etc.}; for this we refer
to the literature \cite{Mueller93,Ji95,MartinelliSachrajda96}.
\par
As was shown in Section \ref{sec_instantons}, the instanton vacuum implies
that among the twist--4 matrix elements appearing in the
$1/Q^2$--corrections to $F_L$ and $F_2$, Eqs.(\ref{B_def}) and
(\ref{C_def}), the one of the quark--gluon operator $B_{\alpha\beta}$ is
parametrically leading and has a large positive value, while those of the
operator $A_{\alpha\beta}$ and the four--fermionic operator
$C_{\alpha\beta}$ are parametrically suppressed. We notice that the
dominant matrix element, $B$, enters with a large (and positive)
coefficient in the twist--4 corrections to $F_L$, Eq.(\ref{M_L_twist4}),
but with a much smaller coefficient in $F_2$,
Eq.(\ref{M_2_twist4}). Furthermore, in the case of $F_2$,
Eq.(\ref{M_2_twist4}), the parametrically suppressed matrix elements $A$
and $C$ have comparatively large coefficients.  In the case of $F_L$ the
situation is more fortunate. Not only does the dominant matrix element in
the instanton vacuum, $B$, enter with a coefficient of order unity, but
also the coefficient of the parametrically suppressed matrix element $A$ is
small. Thus, for $F_L$ we can make a quantitative prediction for the sign
and magnitude of the twist--4 contribution. In addition, for this structure
function the twist--2 contribution is non-zero only in order $\alpha_S$,
{\it cf.}\ Eq.(\ref{M_L_ex_twist2}), so the relative importance of power
corrections is larger than in $F_2$, which is indeed seen in the data for
$R$ \cite{Whitlow90,Bodek98}. We shall therefore concentrate on $F_L$ in
the following.
\par
We would like to compare our theoretical result for the scale dependence of
$\widetilde{F}_L$ with measurements of the ratio $R$ for a deuteron target
\cite{Whitlow90,Bodek98}.  We assume here that $\widetilde{F}_L^{\rm d}
\approx \frac{1}{2} (\widetilde{F}_L^{\rm p} + \widetilde{F}_L^{\rm n} )$,
that the deuteron structure function approximately corresponds to the
isoscalar combination of proton and neutron structure functions.  The full
theoretical result for the $Q^2$--dependence of the second moment of $F_L$
consists of the twist--2, target mass, and dynamical twist--4
contributions, see Section \ref{sec_power}.  We evaluate the twist--2 and
target mass contributions, Eqs.(\ref{M_L_ex_twist2}) and
(\ref{M_L_ex_target}), for the isoscalar target, using the GRV94 NLO
parametrization\footnote{The use of a NLO parametrization for the twist--2
parton distributions is consistent with the use of the
$O(\alpha_S)$--expressions for the Wilson coefficients in the twist--2
contribution.}; see Ref.\cite{GRV95} for details such as the number of
quark flavors, $\Lambda_{\rm QCD}$, {\it etc}.  For the dynamical twist--4
contribution we take the values obtained from the instanton vacuum,
Eq.(\ref{B_final}), and neglect the contribution of strange quarks in the
twist--4 corrections. This results in a total twist--4 contribution of
\be
\widetilde{M}_L^{\rm d} (2, Q^2)^{\rm twist-4} &\approx& 
\frac{3}{4 Q^2} \times \frac{1}{2} (e_u^2 + e_d^2 ) B_S
\;\; = \;\; \frac{(260 \, {\rm MeV})^2}{Q^2} .
\label{twist4_inst}
\ee
\par
\begin{figure}
\setlength{\epsfxsize}{15cm}
\setlength{\epsfysize}{15cm}
\epsffile{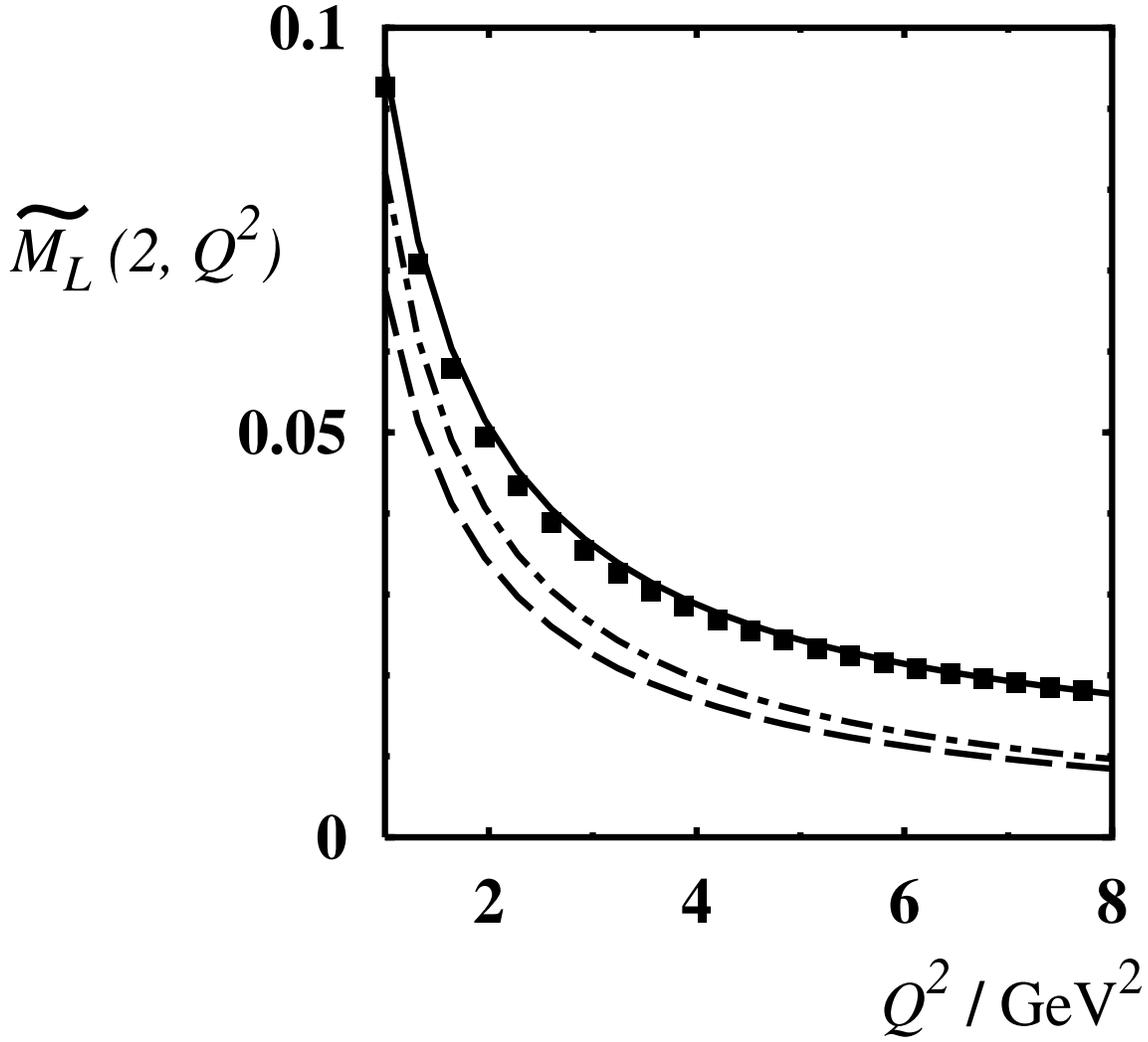}
\caption[]{The $Q^2$--dependence of the second moment of $\widetilde{F}_L$
for the deuteron, $\widetilde{M}^{\rm d}_L (2, Q^2 )$,
Eq.(\ref{moments_def}). {\it Dashed line:} Dynamical twist--4 contribution
obtained from the instanton vacuum, Eq.(\ref{twist4_inst}).  {\it
Dot--dashed line:} Twist--4 contribution plus target mass corrections,
Eq.(\ref{M_L_ex_target}). {\it Solid line:} Total theoretical result,
including also the twist--2 radiative corrections,
Eq.(\ref{M_L_ex_twist2}).  {\it Squares:} ``Experimental'' result for
$\widetilde{M}_L (2, Q^2 )$ obtained by combining the fit of $F_2^{\rm d}$
of Ref.\cite{NMC95} with the fit of $R^{\rm d}$ of Ref.\cite{Whitlow90},
keeping only terms up to order $1/Q^2$ (see the text for details).}
\label{fig_ml}
\end{figure}

It is instructive to first compare the relative magnitude of the power
corrections and the $O(\alpha_S )$--perturbative contributions to
$\widetilde{M}_L$, see Fig.\ref{fig_ml}. The dashed line shows the
dynamical twist--4 contributions obtained from the instanton vacuum,
Eq.(\ref{twist4_inst}). The dot--dashed line is the sum of twist--4 and
target mass corrections, Eq.(\ref{M_L_ex_target}), {\it i.e.}, the total
$1/Q^2$ power corrections. The solid line, finally, is the total result, in
which also the twist--2 contribution has been added; the variation of the
twist--2 contribution with $Q^2$ over the interval shown is much weaker
than that of the power corrections.  One sees that the total power
correction is sizable, which is clearly a consequence of the missing
tree--level [$O(\alpha_S^0 )$--] perturbative contribution, and that power
corrections are dominated by the dynamical twist--4 contribution, which is
much larger than the target mass corrections. These facts combined make the
longitudinal structure function an excellent tool for extracting
quantitative information about higher--twist matrix elements.
\par
We also compare our total theoretical result for $\widetilde{M}_L^{\rm d}$
to order $1/Q^2$ with the moment extracted from the fit to the SLAC data
for $R^{\rm d}$ \cite{Whitlow90}, which describes well also the more recent
deuteron data from the NMC experiment \cite{NMC97}. In order to obtain an
estimate for isosinglet $\widetilde{F}_L^{\rm d}$ to accuracy $1/Q^2$ we
have extracted from the fit for $R^{\rm d}$ in Ref.\cite{Whitlow90} the
total contribution up to order $1/Q^2$, and combined this with the fit to
the NMC data for $F_2^{\rm d}$ \cite{NMC95}, which also includes
$1/Q^2$--corrections. Substituting the fits for $R^{\rm d}$ and $F_2^{\rm
d}$ in Eq.(\ref{F_L_ex_from_R}), retaining only terms up to order $1/Q^2$,
we obtain a rough estimate of $\widetilde{F}_L$ to order $1/Q^2$. To be
able to compute the moment we have estimated the contribution of the
unmeasured small--$x$ region using the GRV parametrization; this
contribution turns out to be at the few percent level. The ``experimental''
result for the moment thus obtained is shown by the dotted line in
Fig.\ref{fig_ml}; it should be compared to the solid line giving the total
theoretical result.  As can be seen the agreement between the total results
is rather satisfactory. This indicates that the twist--4 contribution
derived from instantons is of the right order of magnitude.
\par
We note that our results for the twist--4 matrix elements obtained from the
instanton vacuum are consistent with those of the analysis of power
corrections to the second moments of $F_L$ and $F_2$ of Ref.\cite{Choi93},
which was based on NMC and SLAC data.  Assuming only that the matrix
elements of the four--fermionic operators, $A$ and $C$, Eq.(\ref{A_def})
and (\ref{C_def}), are not strongly different in magnitude, the authors of
Ref.\cite{Choi93} concluded that it is likely that the matrix element $B_S$
has a large positive value of the order of $(380 \pm 95\,{\rm MeV})^2$ at
$\mu^2 = 5 \,{\rm GeV}^2$, somewhat smaller than our
estimate.\footnote{Here we have translated the results of Ref.\cite{Choi93}
into our conventions.} We have repeated the analysis of Ref.\cite{Choi93}
using the results of a recent re-analysis of SLAC and NMC data of
Ref.\cite{Bodek98}, where the twist--4 corrections were parametrized using
a renormalon fit, and found only small changes to the values quoted in
Ref.\cite{Choi93}. In particular, as explained in Subsection
\ref{subsec_effective_operators}, the instanton vacuum provides a
parametric explanation why the contributions of the four--fermionic
operators are suppressed.
\par
\begin{figure}
\setlength{\epsfxsize}{15cm}
\setlength{\epsfysize}{15cm}
\epsffile{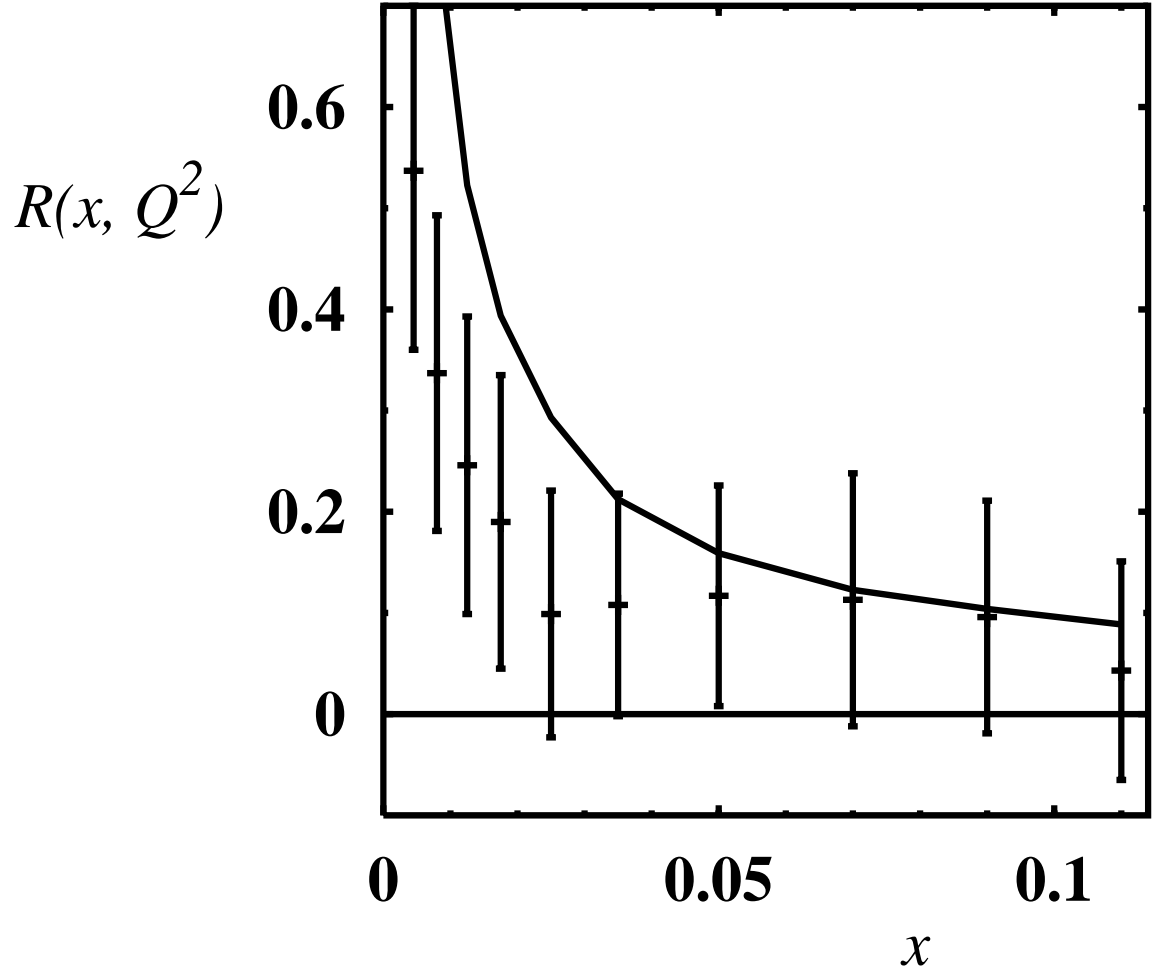}
\caption[]{The ratio $R^{\rm d} = \sigma^{\rm d}_L / \sigma^{\rm d}_T$
obtained from the renormalon model ({\it solid line}), compared with the
NMC data from Ref.\cite{NMC97}. Here the $x$--dependence of the
$1/Q^2$--contribution is predicted by the renormalon formula of
Refs.\cite{MSSM97,DW96}, while the coefficient is determined from the
instanton vacuum result, Eq.(\ref{twist4_inst}).  In the data set the
values of $Q^2$ vary with $x$; see Ref.~\cite{NMC97} for details. Note that
the simple non-singlet renormalon formula is not applicable at small $x$.}
\label{fig_renorm}
\end{figure}

A more conclusive comparison with phenomenological power corrections could
be made if we had results for the $x$--dependence of the twist--4
contribution. In the OPE approach of Refs.\cite{SV82,JS82} this would
require the computation of the nucleon matrix elements of higher--spin
operators describing power corrections to higher moments, a problem which
is currently under investigation. In contrast, in the renormalon approach
to power corrections \cite{Stein96,MSSM97,DW96} the $x$--dependence of the
$1/Q^2$--correction is obtained with relative ease, while it is impossible
to predict the absolute normalization. Fits to the experimental data for
$R$ and $F_2$ indicate that the renormalon formulas describe the
$x$--dependence of the data rather well, at least for large $x$
\cite{Bodek98,RSB99}. In this situation it is interesting to make a rough
estimate of the full $x$--dependent twist--4 contribution simply combining
the two approaches, {\it i.e.}, taking the $x$--dependence from the
renormalon formula and determining the coefficient by the instanton vacuum
calculation. We use the renormalon expression for the twist--4 correction
to $F_L$ given in Refs.\cite{Stein96,DW96}, including only the non-singlet
coefficients, which we evaluate using the GRV94 NLO parametrization for the
twist--2 parton distributions. (The logarithmic $Q^2$--dependence of the
twist--2 distribution may be neglected in this context.) Combining this
with the twist--2 plus renormalon fit to $F_2$ of Ref.\cite{Bodek98} we
obtain an estimate for $R(x, Q^2)$ to order $1/Q^2$, shown by the solid
line in Fig.\ref{fig_renorm}. It is compared to the recent NMC data for
$R(x, Q^2)$ \cite{NMC97}, taken at various $Q^2$. In view of the simplicity
of this model the agreement is quite reasonable. Note that this model is
not applicable in the small--$x$ region.
\par
Finally, in Table \ref{tab1} we compare the instanton vacuum result for the
total twist--4 contribution to the second moment of $F_L$,
Eq.(\ref{twist4_inst}), with results of other approaches.  In the analysis
of Ref.\cite{Bodek98} the higher--twist contributions to $R$ and $F_2$ at
large $x$ were modeled by the non-singlet renormalon contribution
\cite{Stein98,MSSM97,DW96}, with the coefficients treated as a fitting
parameter. Using these fits to construct $F_L$, and computing the second
moment, we obtain the value quoted in Table \ref{tab1}. Also given is the
result of a recent similar renormalon analysis \cite{RSB99}. Both values
are in reasonable agreement with the instanton calculation. Finally, within
a given renormalization scheme it is possible to determine the absolute
magnitude (but not the sign) of the renormalon contribution; we give in
Table \ref{tab1} the value obtained in Ref.\cite{Stein98} in the $\rm
\overline {MS}$ scheme.
%
%
\begin{table}
\begin{tabular}{|l|l|c|}
\hline
Source     & $\kappa^2 / {\rm GeV^2}$  & Scale / ${\rm GeV^2}$ \\
\hline
Present work (instanton calculation)  & $\;\;\; 0.068$ & $\sim 0.36$ 
\\ 
Choe {\it et al.}\ (data analysis)
\cite{Choi93} & $\;\;\; 0.030 \pm 0.015$ & 5
\\
Yang and Bodek (renormalon fit) \cite{Bodek98} &
$\;\;\; 0.034 \pm 0.002$ & 1 
\\
Ricco {\it et al.}\ (renormalon fit) \cite{RSB99} & 
$\;\;\; 0.060 \pm 0.020$ & 1 
\\
Stein {\it et al.} ($\rm \overline {MS}$ renormalon model)
\cite{Stein96,Stein98} & $\pm 0.032$ & 1 
\\
\hline
\end{tabular}
\caption[]{The total twist--4 contribution to the second moment of 
$\widetilde F_L$, parametrized here as 
$\widetilde M_L (2, Q^2)^{\rm twist-4} = \kappa^2/Q^2$,
as extracted from various fits to experimental data or model 
calculations.}
\label{tab1}
\end{table} 
\par
In the so-called transverse basis of twist--4 operators the twist--4
contribution to $F_L$ can be related to the ``average transverse momentum''
of the quarks in the nucleon \cite{EFP82}. Forms such as
\be
\widetilde{M}_L^{\rm d} (2, Q^2)^{\rm twist-4} 
&=& \frac{1}{2} (e_u^2 + e_d^2 ) \frac{4 \langle k_T^2 \rangle}{Q^2}
\label{F_L_from_k_perp}
\ee
are also frequently used as a phenomenological parametrization of the
twist--4 contribution to $F_L$. Comparing Eq.(\ref{F_L_from_k_perp}) with
the OPE result in the ``longitudinal'' basis, Eq.(\ref{M_L_twist4}), and
keeping only the dominant matrix element in the instanton vacuum, $B_S$, we
see that our result for $B_S$, Eq.(\ref{B_final}), corresponds to
\be
\langle k_T^2 \rangle &=& \frac{3}{16} B_S 
\;\; \approx \;\; (250 \, {\rm MeV})^2 .
\label{k_perp_from_instantons}
\ee
It is interesting that $\langle k_T^2 \rangle$ is parametrically of order
$\bar\rho^{-2}$, {\it i.e.}, it is determined by the instanton size, or, in
other words, the ``size'' of the constituent quark, not by the size of the
nucleon, which is of order $M^{-1}$. Thus, the smallness of the instantons
explains the relatively large value of $\langle k_T^2 \rangle$ found in
power corrections.
\section{Conclusions}
\setcounter{equation}{0}
\label{sec_conclusions}
In this paper we have studied the twist--4 matrix elements describing
$1/Q^2$--corrections to unpolarized structure functions in the instanton
vacuum. We have shown that the dominant twist--4 matrix elements are those
of the ``true'' quark--gluon operators, {i.e.}, where the gluon field
cannot be eliminated using the equations of motion, not those of the
four--fermionic (or ``diquark'') operators.  The typical scale for the
matrix elements of the former is the square of the instanton size,
$\bar\rho^{-2}$, while the latter are of order of the dynamical quark mass
squared, $M^2$, and thus suppressed by the instanton packing fraction,
$(\bar\rho / \bar R)^4$. This result is contrary to the widely held belief,
based on a naive ``constituent quark'' picture, that it is the diquark
correlations which are mostly responsible for power corrections to
structure functions. Rather, in our picture, the dominant power corrections
come from the interaction of the quarks with non-perturbative vacuum
fluctuations of the gauge fields of small size.
\par
The numerical values obtained for the various twist--4 matrix elements
follow their parametric ordering. The dominant quark--gluon operator,
$B_{\alpha\beta}$, has a large positive value, while the matrix elements of
the diquark operators are an order of magnitude smaller. This seems to be
in qualitative agreement with the results of a phenomenological analysis of
power corrections to $F_2^{\rm d}$ and $R^{\rm d}$. We are not claiming, of
course, that the observed power corrections could only be explained by
instantons.  Contrary to the phenomenon of chiral symmetry breaking, which
can directly be linked to the presence of topologically non-trivial vacuum
fluctuations, in the case of power corrections to structure functions there
is no {\it a priori} reason why instantons should play a role.
Nevertheless, it seems that the basic assumptions made in the instanton
vacuum, namely the dominance of instanton--type fluctuations and diluteness
of the instanton medium, are consistent with the information we have on
power corrections to structure functions.
\par
We have shown that in the instanton vacuum the effective operator for the
leading quark--gluon operator $B_{\alpha\beta}$ can be reduced to a
one--body quark operator measuring the ``average virtuality'' of the quarks
in the nucleon.  The sign and magnitude of the twist--4 matrix element $B$
depend not only on our picture of non-perturbative vacuum fluctuations, but
also on our description of the nucleon. In this respect it is crucial that
the picture of the nucleon as a chiral soliton, which is a fully
relativistic and field--theoretic treatment, is obtained from the effective
low--energy theory without making any additional approximations. The
correct sign of the matrix element $B_S$ is obtained in this approach due
to the fact that the average momentum of the quarks in the nucleon is
space-like, which would not be the case, say, in a simple constituent quark
model.
\par
The approach taken in the present work can be extended in many ways. A
straightforward continuation would be the computation of the matrix
elements of twist--4 operators giving power corrections to higher moments
of the structure functions, and the restoration of the $x$--dependence of
the twist--4 contribution. It is likely that twist--4 QCD operators of spin
$>2$ can again be represented in the effective theory by simple one--body
quark operators of the form of twist--2 operators with extra derivatives,
as in the case of spin $2$, Eq.(\ref{B_simple}).  A much more challenging
task would be to depart from the notion of power corrections altogether and
develop a uniform description of the $Q^2$--dependence of structure
functions, valid also at small $Q^2$, {\it e.g.}\ in the spirit of
Ref.\cite{BKS97}. The instanton vacuum, which on one hand has a clear
relation to perturbative QCD, on the other hand describes non-perturbative
phenomena in the infrared domain, may be a good starting point for such
attempts.
\par
Finally, we note that the methods developed here can be applied to study
also power corrections to exclusive processes such as deeply--virtual
Compton scattering and hard meson production, which may be of great
practical importance \cite{VGG99}. For these processes, however, up to now
no systematic classification of power corrections in QCD is available.
\\[.5cm]
{\large\bf Acknowledgements} \\[.2cm]
We are grateful to M.V.\ Polyakov for many valuable suggestions throughout
the course of this work, and to D.I.\ Diakonov, K.\ Goeke, Nam--Young Lee,
L.\ Mankiewicz, V.Yu.\ Petrov, P.V.\ Pobylitsa, A.\ Sch\"afer, and E. Stein
for numerous interesting discussions.
\\[.2cm]
This work has been supported in part by the Deutsche Forschungsgemeinschaft
(DFG), by the German Ministry of Education and Research (BMBF), and by COSY
(J\"ulich). M.M.\ acknowledges support from the TMR Programme of the
European Commission.
\newpage
\appendix
\renewcommand{\theequation}{\Alph{section}.\arabic{equation}}
\section{Operator product expansion for unpolarized DIS}
\setcounter{equation}{0}
\label{app_ope}
In this appendix we present a short derivation of the QCD predictions for
the $Q^2$--de\-pen\-dence of the second moments of $F_2$ and $F_L$ using
the operator product expansion.
\par
From the definition of the hadronic tensor, Eq.(\ref{W_def}), explicit
expressions for the structure functions $F_2$ and $F_L$ can be obtained by
contracting Eq.(\ref{W_def}) with $p^\mu p^\nu$, taking its trace, and
solving the resulting equations:
\be
F_L (x, Q^2) &=& \frac{4 x^3}{Q^2} \left( M_N^2 W + 2 W_{pp} \right) 
\; + \; O\left( \frac{1}{Q^4} \right) ,
\label{F_L_from_W}
\\
F_2 (x, Q^2) &=& - x W \; + \; \frac{4 x^3}{Q^2} 
\left( M_N^2 W + 3 W_{pp} \right) 
\; + \; O\left( \frac{1}{Q^4} \right) ,
\label{F_2_from_W}
\ee
where
\be
W &\equiv& W^\mu_\mu , \hspace{2cm} 
W_{pp} \;\; \equiv \;\; p^\mu p^\nu W_{\mu\nu} .
\label{W_pp_def}
\ee
We have kept only terms up to order $1/Q^2$. Note that $W$ and $W_{pp}$
generally contain terms of any order in $1/Q^2$ and should be substituted
in Eqs.(\ref{F_L_from_W}, \ref{F_2_from_W}) consistently.
\par
The QCD description makes use of the fact that, by the optical theorem, the
hadronic tensor is related to the imaginary part of the forward virtual
Compton amplitude,
\be
T_{\mu\nu} (p, q) &=& i \int d^4 z \; e^{i q\cdot z}
\langle p | T\left\{ J_\mu (z) J_\nu (0) \right\} | p \rangle .
\label{T_def}
\ee
As a function of the photon energy, $q^0$, it exhibits a cut in the
inelastic region. To express the hadronic tensor in terms of the Compton
amplitude it is convenient to regard the latter as a function of the
dimensionless variable
\be
\omega &\equiv& \frac{1}{x} \;\; = \;\; \frac{2 q\cdot p}{Q^2} ;
\ee
in this variable the cut runs along the real axis from $-\infty$ to $-1$
and from $+1$ to $+\infty$.  In the case of electro-magnetic scattering the
Compton amplitude is crossing--even, and thus an even function of
$\omega$. The relation to the hadronic tensor is provided by the formula
\be
\left.
\begin{array}{l} 
T_{\omega^n} \\[1ex]
T_{pp, \omega^n} 
\end{array} \right\}
&=& 
4 \int_{0}^1 dx x^{n - 1} 
\left\{
\begin{array}{r} 
W(x) \\[1ex]
W_{pp} (x) 
\end{array} ,
\right.
\label{W_from_t}
\ee
where $T$ and $T_{pp}$ are defined in analogy to Eq.(\ref{W_pp_def}), and
$T_{\omega^n}, T_{pp, \omega^n}$ denote the coefficients of $\omega^n$ in
the power series expansions of $T$ and $T_{pp}$ in $\omega$ in the domain
$|\omega | < 1$. In particular, for the second moments of $F_L$ and $F_2$
one obtains in this way:
\be
\int_0^1 dx \, F_L (x, Q^2 )
&=& \frac{1}{Q^2} \left( M_N^2 T_{\omega^4} + 2 T_{pp, \omega^4} \right)
\; + \; O\left( \frac{1}{Q^4} \right) ,
\label{F_L_from_T}
\\
\int_0^1 dx \, F_2 (x, Q^2 ) 
&=& -\frac{1}{4} T_{\omega^2} \; + \;
\frac{1}{Q^2} \left( M_N^2 T_{\omega^4} + 3 T_{pp, \omega^4} \right)
\; + \; O\left( \frac{1}{Q^4} \right) .
\label{F_2_from_T}
\ee
The general strategy is now clear. One computes the virtual Compton
amplitude in the Bjorken limit, $Q^2 \rightarrow \infty$ and $\omega$
fixed, expanding in $1/Q^2$, and pick up the coefficients in $T$ and
$T_{pp}$ of the given powers in $\omega$.
\par
The dominant contribution to the virtual Compton amplitude in the Bjorken
limit is given by the ``handbag'' diagram describing the scattering of the
photon off a single quark. At level $1/Q^2$ one must consider generally two
types of contribution, corresponding to the different contractions of the
quark fields in Eq.(\ref{T_def}) shown in
Fig.\ref{fig_twist4}. Contributions of type (b) can be computed using
standard Feynman diagrams and give rise to matrix elements of
four--fermionic operators in the nucleon (see below). Contributions of type
(a) require the calculation of the quark Green function in the background
of the non-perturbative gluon field of the nucleon. This can conveniently
be done using the Schwinger method \cite{SV82,Schwinger51}. In this
approach the contribution (a) to the Compton amplitude is given by the
formal expression ($\hat q \equiv q^\alpha \gamma_\alpha$ {\it etc.})
\be
T_{\mu\nu} &=& \sum_f e_f^2 \; \langle p|  
\bar\psi_f \left[ - \gamma_\mu \frac{1}{\hat P + \hat q} 
\gamma_\nu
- \gamma_\nu \frac{1}{\hat P - \hat q} \gamma_\mu \right] 
\psi_f |p \rangle ,
\label{T_short}
\ee
where
\be
P_\mu &\equiv& i \nabla_\mu \;\;\; = \;\;\; i \partial_\mu + A_\mu , 
\hspace{2cm}
[P_\mu , P_\nu ] \;\; = \;\; i F_{\mu\nu} .
\ee
Eq.(\ref{T_short}) is understood as a starting point for expanding in 
local operators:
\be
\frac{1}{\hat P \pm \hat q} &=& 
\pm \frac{\hat q}{q^2} 
- \frac{\hat q \hat P \hat q}{(q^2 )^2}  
\pm \frac{\hat q \hat P \hat q \hat P \hat q}{(q^2 )^3} 
- \frac{\hat q \hat P \hat q \hat P \hat q \hat P 
\hat q}{(q^2 )^4} 
+ \ldots 
\label{Schwinger_expansion}
\ee
To order $1/Q^2$ all contributions to the second moments of $F_L$ and
$F_2$, Eqs.(\ref{F_L_from_T}) and (\ref{F_2_from_T}), are contained in the
$O(P)$ and $O(P^3 )$ terms of Eq.(\ref{Schwinger_expansion}). We now list
these contributions.  For simplicity we suppress the sum over flavors and
the quark charges; they will be inserted again at the end of the
calculation.
\par
The $O(P)$--term of Eq.(\ref{Schwinger_expansion}) contributes to
$T_{\omega^2}$ and gives rise to the scaling part of the second moment of
$F_2$:
\be
T (\mbox{order $P$})
&\equiv& T_\mu^{\;\;\mu}(\mbox{order $P$}) \;\; = \;\; 
\frac{2}{(q^2 )^2} \langle p|  
\bar\psi \gamma^\mu \hat{q} \hat P \hat{q} \gamma_\mu \psi
|p \rangle .
\ee
Using the identity
\be
\hat{q} \hat P &=& 2 q\cdot P - \hat P \hat{q},
\label{qPq}
\ee
as well as $\gamma^\mu \hat{q} \gamma_\mu = -2\hat{q}$, one can write this
as
\be
T(\mbox{order $P$}) &=& \;\; = \;\; -\frac{8}{(q^2 )^2} q^\alpha q^\beta
\langle p|  \bar\psi P_\alpha \gamma_\beta \psi |p \rangle 
\; + \; \ldots
\label{T_1}
\ee
where we have dropped terms with contracted $q$'s which do not contribute
to the $\omega^2$--term. We see that this contribution is entirely given by
the nucleon matrix element of the spin--2 twist--2 operator,
\be
\langle p|  \bar\psi P_\alpha \gamma_\beta \psi
|p \rangle &=& 2 L^{(2)} (p_\alpha p_\beta - \textfrac{1}{4} p^2 
g_{\alpha\beta}) ;
\label{spin_2}
\ee
the traceless structure here is a consequence of the QCD equations of
motion: $\bar\psi \hat P \psi = 0$. Substituting Eq.(\ref{spin_2}) in
Eq.(\ref{T_1}) one finds
\be
T(\mbox{order $P$}) &=& -4 L^{(2)} \omega^2 \; + \; \ldots ,
\ee
and from Eq.(\ref{F_2_from_T}) one obtains
\be
\int_0^1 dx\; F_2 (x) &=& L^{(2)} + O\left( \frac{1}{Q^2} \right) .
\ee
\par
The $O(P^3)$--term in the expansion Eq.(\ref{Schwinger_expansion})
contributes to the $\omega^2$ as well as the $\omega^4$--part of $T$, and
to the $\omega^4$--part of $T_{pp}$. It gives rise to the dynamical
twist--4 as well as the target mass corrections to the second moments. One
can easily see that the dynamical twist--4 contributions are contained in
$T_{\omega^2}$ and $T_{pp, \omega^4}$, since a twist--4 matrix element can
contribute at most two powers of the target momentum. (In particular, this
means that the twist--4 contribution to the second moment of $F_L$ comes
only from $T_{pp, \omega^4}$.) Target mass corrections, on the other hand,
arise from all terms. We begin with
\be
T_{pp}(\mbox{order $P^3$}) &=& \frac{2}{(q^2 )^4} \, p^\mu p^\nu \,
\langle p|  \bar\psi \gamma_\mu \hat{q} \hat P \hat{q} 
\hat P \hat{q} 
\hat P \hat{q} \gamma_\nu \psi |p \rangle .
\ee
Making repeated use of the identity Eq.(\ref{qPq}), dropping along the way
all terms with $\hat{q}\hat{q} = q^2$ in the numerator which do not
contribute to the $\omega^4$--piece, one can reduce this to
\be
\lefteqn{
T_{pp}(\mbox{order $P^3$}) \;\; = \;\; \frac{2}{(q^2 )^4} \, p^\mu p^\nu \,
\langle p|  \bar\psi (2 q\cdot P)^3 \gamma_\mu \hat{q} \gamma_\nu \psi 
|p \rangle \; + \; \ldots } &&
\nonumber \\
&=& \frac{16}{(q^2 )^4}
\left[ (2q\cdot p) q^\alpha q^\beta q^\gamma p^\delta
- M_N^2 q^\alpha q^\beta q^\gamma q^\delta \right]
\; \langle p|  \bar\psi P_\alpha P_\beta P_\gamma 
\gamma_\delta \psi |p \rangle \; + \; \ldots
\label{T3_pp_res}
\ee
The reduction of the trace term, $T$, is slightly more complicated, since
one has to retain the $q^2$--terms, which can contribute to the
$\omega^2$--piece. Again using Eq.(\ref{qPq}) one first obtains
\be
T(\mbox{order $P^3$}) &=& \frac{2}{(q^2 )^4} 
\langle p|  \bar\psi \gamma^\mu \hat{q} \hat P \hat{q} 
\hat P \hat{q} 
\hat P \hat{q} \gamma_\mu \psi |p \rangle 
\nonumber \\
&=& \frac{2}{(q^2 )^4} \langle p| \left\{
(2q\cdot P)^3 \gamma^\mu \hat{q} \gamma_\mu
- q^2 \left[ (2q\cdot P) \gamma^\mu \hat P \gamma_\mu 
(2q\cdot P) \right. \right. 
\nonumber \\ 
&& \left. \left. 
\;\;\;\;\; + (2q\cdot P) \gamma^\mu \hat{q} \hat P 
\hat P \gamma_\mu 
+ \gamma^\mu \hat P \hat P \hat{q} \gamma_\mu 
(2q\cdot P) 
\right] 
\right\} |p \rangle + \ldots
\ee
Anticommuting now the matrices $\gamma_\mu$ from the left to the right
position, and taking note of the QCD equations of motion
\be
\bar\psi \, \ldots \hat P \, \psi
&=& 
\bar\psi \, \hat P \ldots \, \psi \;\; = \;\; 0,
\ee
one finally obtains
\be
T(\mbox{order $P^3$}) &=& \frac{16}{(q^2 )^4} \left[
-2 q^\alpha q^\beta q^\gamma q^\delta
+ q^2 \left( - q^\alpha q^\gamma g^{\beta\delta}
+ q^\alpha q^\delta g^{\beta\gamma}
+ q^\gamma q^\delta g^{\alpha\beta} \right)
\right]
\nonumber \\
&& \times
\langle p|  \bar\psi P_\alpha P_\beta P_\gamma 
\gamma_\delta \psi
|p \rangle \; + \; \ldots
\label{T3_res}
\ee
\par
We see that both Eqs.(\ref{T3_pp_res}) and (\ref{T3_res}) are determined by
the matrix element of the reducible rank--4 tensor operator, 
$\bar\psi P_\alpha P_\beta P_\gamma \gamma_\delta \psi$.  Only the part
symmetric in the indices $\alpha$ and $\gamma$ enters. On general grounds,
this matrix element can be parameterized as\footnote{On grounds of Lorentz
invariance and symmetry in $\alpha, \gamma$ Eq.(\ref{ansatz_symm}) could
contain also a structure 
$i (\epsilon_{\sigma\beta\gamma\delta} S_{\sigma\alpha} +
\epsilon_{\sigma\beta\alpha\delta} S_{\sigma\gamma})$, which, however, is
not related to the matrix element of a twist--4 operator.}
\be
\langle p|  \bar\psi P_{\left\{ \alpha \right\} } P_\beta 
P_{\left\{ \gamma \right\} } \gamma_\delta \psi
|p \rangle &=& 2 L^{(4)} S_{\alpha\beta\gamma\delta}
\nonumber \\
&+& 2 b_1 
\left( g_{\alpha\beta} S_{\gamma\delta} 
+ g_{\gamma\beta} S_{\alpha\delta} \right)
+ 2 b_2 g_{\alpha\gamma} S_{\beta\delta} 
\nonumber \\
&+& 2 b_3
\left( g_{\alpha\delta} S_{\beta\gamma} 
+ g_{\gamma\delta} S_{\alpha\beta} 
\right)
+ 2 b_4 g_{\beta\delta} S_{\alpha\gamma} 
\nonumber \\
&+& \mbox{terms $g_{\alpha\beta} g_{\gamma\delta}$ etc.} 
\label{ansatz_symm}
\ee
Here $S_{\alpha\beta\ldots}$ denotes the totally symmetric traceless tensor
constructed from the vector $p_\mu$:
\be
S_{\alpha\beta\gamma\delta}
&=& p_\alpha p_\beta p_\gamma p_\delta
- \textfrac{1}{8} p^2 \left( 
g_{\alpha\beta} p_\gamma p_\delta + \ldots +
g_{\gamma\delta} p_\alpha p_\beta \right)
\nonumber \\
&& + \textfrac{1}{48} p^4 \left( g_{\alpha\beta} g_{\gamma\delta} 
+ g_{\alpha\gamma} g_{\beta\delta} 
+ g_{\alpha\delta} g_{\beta\gamma} 
\right) , 
\\
S_{\alpha\beta} &=& p_\alpha p_\beta 
- \textfrac{1}{4} p^2 g_{\alpha\beta} .
\ee
The coefficients in the ansatz Eq.(\ref{ansatz_symm}) can be expressed in
terms of matrix elements of operators of definite spin (and
twist). $L^{(4)}$ is determined by the matrix element of the unique
spin--4, twist--2 operator
\be
\langle p|  \bar\psi P_{\left\{ \alpha \right. } P_\beta 
P_\gamma
\gamma_{\left. \delta \right\} } \psi
|p \rangle - \mbox{traces}
&=& 2 L^{(4)} S_{\alpha\beta\gamma\delta} .
\label{spin_4}
\ee
When inserted in Eqs.(\ref{T3_pp_res}) and (\ref{T3_res}) this part of the
matrix element gives contributions
\be
T(\mbox{order $P^3$}) &=& 
L^{(4)} \left( -4 \omega^4 - \frac{12 M_N^2}{Q^2} \omega^2 \right) ,
\\
T_{pp}(\mbox{order $P^3$}) &=& \frac{M_N^2}{2} L^{(4)} \omega^4 ,
\ee
and thus gives rise to the ``target mass'' corrections\footnote{This result
agrees with Ref.\cite{SV82} when one expands the Nachtmann moments there in
$M_N^2/Q^2$.}
\be
\int_0^1 dx\, F_2 (x, Q^2 )^{\rm target\; mass} 
&=& \frac{L^{(4)}}{2} \frac{M_N^2}{Q^2} ,
\\
\int_0^1 dx\, F_L (x, Q^2 )^{\rm target\; mass} 
&=& -3 L^{(4)} \frac{M_N^2}{Q^2} .
\ee
The coefficients $b_1 \ldots b_4$ in Eq.(\ref{ansatz_symm}) can be
expressed in terms of the matrix elements of certain spin--2 twist--4
operators related to contractions of the rank--4 operator. We consider the
set of operators
\be
A_{\alpha\beta} &=& \bar\psi \gamma_\alpha
\left[\nabla^\gamma , F_{\gamma\beta} \right] \psi
\;\; = \;\; \bar\psi \gamma_\alpha (D^\gamma F_{\gamma\beta}) \psi ,
\\[1ex]
B_{\alpha\beta} &=& \bar\psi \gamma^\gamma \gamma_5 \, i 
\left( \nabla_\alpha \Fdual_{\beta\gamma} 
+ \Fdual_{\beta\gamma} \nabla_\alpha \right) \psi ,
\\[1ex]
D_{\alpha\beta} &=& \bar\psi \gamma^\gamma 
\left[ \nabla_\alpha , F_{\beta\gamma} \right] \psi
\;\; = \;\; \bar\psi \gamma^\gamma (D_\alpha F_{\beta\gamma}) \psi ,
\ee
where symmetrization in $\alpha , \beta$ and subtraction of traces is
implied. Their matrix elements are given by
\be
\langle p| A_{\alpha\beta} |p \rangle &=& 2 A \; S_{\alpha\beta}
\hspace{2cm} \mbox{etc.}
\ee
The expressions for $b_1 \ldots b_4$ can be found by taking traces of both
sides of Eqs.(\ref{ansatz_symm}).  By straightforward calculation one finds
\be
\left.
\begin{array}{l}
g^{\alpha\beta} \\[1ex]
g^{\alpha\gamma} \\[1ex]
g^{\alpha\delta} \\[1ex]
g^{\beta\delta}
\end{array} \right\}
\langle p | 
\bar\psi P_{\{\alpha \}} P_\beta P_{\{\gamma \}} 
\gamma_\delta \psi
| p \rangle
&=&
\left\{
\begin{array}{l}
 ( D - B ) \, S_{\gamma\delta} , \\[1ex]
 ( D - B + A) S_{\beta\delta} , \\[1ex]
0 , \\[1ex]
D\, S_{\alpha\beta} .
\end{array} \right.
\ee
The operator $B_{\alpha\beta}$ comes into play when writing 
$P^\mu P_\mu \, = \, \hat P \hat P - (i/4) [\gamma_\rho , \gamma_\sigma]
F_{\rho\sigma}$ and making use of the three--gamma identity
\be
\gamma_\alpha \gamma_\beta \gamma_\gamma
&=& g_{\alpha\beta} \gamma_\gamma
- g_{\alpha\gamma} \gamma_\beta
+ g_{\beta\gamma} \gamma_\alpha
- i \varepsilon_{\alpha\beta\gamma\delta} \gamma^\delta \gamma_5 .
\ee
We are using the conventions
\be
\gamma_5 &=& -i \gamma^0 \gamma^1 \gamma^2 \gamma^3,
\hspace{1cm} \textfrac{1}{4} {\rm tr} \left[\gamma^\alpha \gamma^\beta 
\gamma^\gamma \gamma^\delta \gamma_5 \right] \;\; = \;\;
i \varepsilon^{\alpha\beta\gamma\delta}, 
\hspace{1cm}
\varepsilon^{0123} \; = \; 1.
\ee
Solving the system of equations obtained by contracting
Eqs.(\ref{ansatz_symm}) for $b_1 \ldots b_4$ one obtains\footnote{This
result is in agreement with that of Ref.\cite{SV82}. Note that in
Ref.\cite{SV82} $t^a \equiv \lambda^a$, and that in Ref.\cite{SV82} the
operator $A_{\mu\nu}$ is defined with different sign and relative
coefficient.}
\be
\left.
\begin{array}{l}
b_1 \\[1ex] 
b_2 \\[1ex]
b_3 \\[1ex]
b_4 \\[1ex]
\end{array}
\right\}
&=& \frac{1}{32}
\left\{
\begin{array}{rrr}
  -A & - 3 B & + 2 D , \\[1ex]
 5 A & - 3 B & + 4 D , \\[1ex]
  -A &   + B & - 2 D , \\[1ex]
   A &   + B & + 4 D . \\[1ex]
\end{array}
\right.
\ee
Inserting the spin--2, twist--4 part of the matrix element in
Eqs.(\ref{T3_pp_res}) and (\ref{T3_res}) one obtains
\be
T(\mbox{order $P^3$}) &=& \frac{1}{Q^2} (A + 5 B) \omega^2 + \ldots ,
\\
T_{pp}(\mbox{order $P^3$}) 
&=& \frac{1}{8} (-A + 3 B) \omega^4 + \ldots ,
\ee
and thus from Eqs.(\ref{F_L_from_T}) and (\ref{F_2_from_T})
\be
\int_0^1 dx \, F_L (x, Q^2 )^{\rm twist-4} 
&=& \frac{1}{4 Q^2} (-A + 3 B), 
\\
\int_0^1 dx \, F_2 (x, Q^2 )^{\rm twist-4} 
&=& \frac{1}{8 Q^2} (-5 A - B) .
\ee
As noted in Ref.\cite{SV82} the operator $D_{\mu\nu}$ does not appear in
the final result for the dynamical twist--4 contributions.
\par
To obtain the final expressions one still has to take into account the
quark charges, {\it i.e.}, replace in all the above matrix elements of
quark bilinear operators
\be
\bar\psi \ldots \psi &\rightarrow& \sum_f e_f^2 \; \bar\psi_f \ldots \psi_f .
\ee
\section{Minkowskian vs.\ Euclidean theory}
\setcounter{equation}{0}
\label{app_eucl}
Here we state our conventions for passing from the Minkowskian to the
Euclidean field theory, in which the instanton vacuum is defined. After
continuing the time dependence of the fields to imaginary times we
introduce Euclidean vector components according to
\be
(x^0 )_{\rm M} &=& -i (x_4 )_{\rm E}, \hspace{1.2em} 
(x^i )_{\rm M} \; = \; (x_i )_{\rm E} .
\label{euclid_x}
\ee
The corresponding rules for derivatives are 
\be
(\partial_0 )_{\rm M} = i (\partial_4 )_{\rm E}, 
\hspace{1em} 
(\partial_i )_{\rm M} = (\partial_i )_{\rm E} ,
\label{euclid_derivative}
\ee
and the Euclidean gauge potential is introduced analogously,
\be
(A_0)_{\rm M} = i (A_4 )_{\rm E} , 
\hspace{1cm}
(A_i )_{\rm M} = (A_i )_{\rm E} ,
\label{euclid_A}
\ee
so that the relative sign between derivative and gauge field in the
covariant derivative, Eq.(\ref{covariant_fundamental_def}), remains the
same in the Euclidean theory. The Euclidean field strength is again defined
by Eq.(\ref{F_from_commutator}). The dual field strength in the Euclidean
theory is defined as $(\widetilde F_{\mu\nu})_{\rm E} = (1/2)
\varepsilon_{\mu\nu\rho\sigma} (F_{\rho\sigma})_{\rm E}$, with
$\varepsilon_{1234} = 1$.
\par
For use in the Euclidean theory we introduce hermitean gamma matrices
according to
\be
(\gamma^0 )_{\rm M} \; = \; (\gamma_4 )_{\rm E}, \hspace{1.2em} 
(\gamma^i )_{\rm M} \; = \; i (\gamma_i )_{\rm E} ,
\label{euclid_gamma}
\ee
and use $\psi^\dagger \equiv i\bar\psi$ for the Dirac conjugate fermion
field (not to be confused with the hermitean conjugate of $\psi$).
%

%
%
\newpage
\newpage
\newpage
\end{document}